\documentclass[english]{revtex4}
\usepackage[T1]{fontenc}
\usepackage{mathrsfs}
\usepackage{amsmath}
\usepackage{amssymb}
\usepackage{esint}

\makeatletter
\@ifundefined{textcolor}{}
{%
 \definecolor{BLACK}{gray}{0}
 \definecolor{WHITE}{gray}{1}
 \definecolor{RED}{rgb}{1,0,0}
 \definecolor{GREEN}{rgb}{0,1,0}
 \definecolor{BLUE}{rgb}{0,0,1}
 \definecolor{CYAN}{cmyk}{1,0,0,0}
 \definecolor{MAGENTA}{cmyk}{0,1,0,0}
 \definecolor{YELLOW}{cmyk}{0,0,1,0}
}

\makeatother

\usepackage{babel}
\begin{document}
\title{Quantum Mechanics Relative to a Quantum Reference System: a Relative
State Approach}
\author{M.J.Luo}
\address{Department of Physics, Jiangsu University, Zhenjiang 212013, People's
Republic of China}
\email{mjluo@ujs.edu.cn}

\begin{abstract}
This paper proposes an intrinsic or background-independent quantum
framework based on entangled state rather than absolute quantum state,
it describes a quantum relative state between the under-study quantum
system and the quantum measuring apparatus as a quantum reference
system, without relying on any external absolute parameter. The paper
focuses on a simple example, in which a quantum object's
one-dimensional position as an under-study quantum system, and a quantum
clock as a quantum reference system or quantum measuring apparatus.
The evolution equation of the state of the quantum object's
position with respect to the state of the quantum clock is given coming
from the Ricci-flat Kähler-Einstein equation. In a linear and
non-relativistic approximation, the framework recovers the equation
of the standard quantum mechanics, in which an intrinsic potential
related to some ``inertial force'' is automatically incorporated
in the covariant derivative. A physical relative probability interpretation
and a geometric non-trivial fiber bundle interpretation of the entangled
state in this intrinsic quantum framework are given. Furthermore,
some non-inertial effects, such as the ``inertial force'', coming
from the general covariance of the intrinsic quantum framework are
also discussed. Compared with the functional integral approach which
is more easily to generalize the quantum clock to the quantum spacetime
reference frame and study quantum gravity, the relative state approach
as a canonical description is more suitable for conceptually demonstrating
the connections to the standard formalism and interpretation of the
quantum mechanics.
\end{abstract}
\maketitle

\section{Introduction}

The formulation and interpretation of quantum theory in its current
form are based on the concept of classical inertial frames. The Schr\"{o}dinger
equation governing the evolution of quantum states in its present
form is valid only in classical inertial frames that employ Newtonian
time as a global absolute parameter. One of the key motivations behind
the development of General Relativity was the absence of the global
absolute time due to the constancy of the speed of light, which necessitated
the removal of the special status of the inertial frames in physics
and the establishment of physical theories (of cause should includes
quantum theory) on concepts that hold true in general coordinate systems.
In geometry, the evolution began from the analytical geometric approach
using a global Cartesian coordinate system, to Gauss's realization
that measuring from local coordinate systems on two-dimensional surface
suffices to define its intrinsic geometry, without the need to embed
it into a three-dimensional space. This was further generalized by
Riemann, who extended Gauss's ideas on intrinsic geometry through
the use of a local quadratic distance form. Starting with Gauss and
culminating in Riemann, the method of intrinsic geometry, independent
of the coordinate systems, was pioneered. This coordinate-independent
spirit or program reached its physical pinnacle in Einstein's classical
General Relativity, bringing about a historic synthesis in classical
physics. However, this spirit of the intrinsic geometry has not been
fully incorporated into quantum theory. Relative to the principles
of the intrinsic geometry and General Relativity, the current framework
of quantum theory still requires and relies on the description in
terms of an external absolute inertial coordinate system. It remains
an ``extrinsic'' theory that necessitates embedding in external
coordinate systems for its description. From this perspective, contemporary
quantum theory can only be seen as a version of ``Newtonian mechanics''
built upon the Cartesian coordinate system. This is also one of the
key reasons for the conflict between the framework of quantum theory
and that of the general principle of relativity. Such an intrinsic,
background-independent quantum framework is particularly urgent for
understanding the quantum properties of gravity based on the intrinsic
spacetime geometry. Quantum theory and General Relativity must be
established on a unified conceptual foundation.

Since the traditional quantum theory necessitates dividing the world
into two distinct parts---a quantum system under-study and an external
absolute classical apparatus or classical ``observer''---it requires
the introduction of several highly controversial assumptions to provide
a coherent explanation of the contemporary quantum mechanics. For
instance, according to the standard Copenhagen interpretation, a quantum
wavefunction with a large spatial extent and extension must undergo
an instantaneous ``superluminal'' collapse across two distant spatial
coordinates (as exemplified by the so-called EPR paradox). Einstein
used such a paradox to express his perceived intrinsic incompatibility
between quantum theory and the relativity. Current technological advancements
have enabled the conduct of long-distance quantum mechanics experiments
spanning several kilometers or even hundreds of kilometers (such as
experiments involving the distribution and measurement of entangled
photons at long distances apart). Although the correlation measurement
results of the entangled photon pairs agree to the predictions of
the contemporary quantum mechanics, the issue does not lie with the
mathematical formalism of quantum mechanics itself, but rather with
its physical interpretation. Specifically, it concerns how an observer
at one spatial location can locally prepare experimental apparatus
to measure a local quantum state and how to appropriately interpret
the ``relative relationship'' (rather than the two individual quantum
states themselves) between this measured quantum state and a quantum
state at another distant spatial location. Currently, there is no
superior interpretation of the mathematical formalism of the contemporary
quantum mechanics that provides a self-consistent and unified framework
for understanding these inherent contradictions. The Copenhagen interpretation
remains the closest to a ``laboratory-common-sense'' and a most
convenient explanation for the mathematical formalism of contemporary
quantum theory, leading to its widespread acceptance among physicists
working at laboratory scales. However, we recognize that when certain
interpretations (through thought experiments) are pushed to extreme
scenarios, they may lead to inherent problems.

Due to the aforementioned shortcomings of the contemporary quantum
mechanics, this paper attempts to establish a quantum mechanical framework
that is (1) intrinsic, background-independent, and geometrized---that
is, it relies solely on the description of internal physical degrees
of freedom without resorting to any external absolute parameters;
(2) treating both the quantum system under-study and the quantum measuring
apparatus on an equal footing; and (3) formulates a theory that describes
the ``interrelationships'' between quantum states, particularly
the relationships between the quantum state of the system under-study
and that of the measuring quantum apparatus, rather than focusing
on individual quantum states relative to an external absolute classical
apparatus. In other words, it employs relative or conditional probabilities
to replace the standard absolute probabilities. In this paper, we
will observe that entangled states provide an appropriate conceptual
foundation for ``internalizing'' or ``relationalizing'' the interpretation
of quantum mechanics, rather than serving as absolute quantum states
in the framework of the contemporary quantum mechanics. This approach
thereby avoids many of the perplexing issues arising from the so-called
``external'' interpretations of the standard Copenhagen interpretation.

Because the vast body of literature on the subject on the relational
formulations and internal evolution of quantum mechanics, it is hardly
for us to fully grasp all the existing works to review here, we only
provide a potentially incomplete overview of some key landmark studies
that we consider crucial and relevant to the topic of this paper,
and how our works differ from these studies are leaving in the section-VII.
The first internal clock time evolution formalism of quantum mechanics
can be attributed to the seminal work of Page and Wootters \citep{PhysRevD.27.2885}
in 1980s. And there are also a large subsequent literature on the
Page-Wootters mechanism, relational time and quantum clock, see e.g.
\citep{Foti:2020erm,Woods2019pagewootters,qfns-48vq,Calcinari:2024pek}
and references therein. The concept of quantum reference frame is
also a long-standing research in literature, including early foundational
works \citep{PhysRevD.30.368} and extensive developments on superselection
rules \citep{RevModPhys.79.555}, relational observables \citep{Rovelli:1990pi},
frame transformations \citep{2013Changing,Giacomini:2017zju,Lake:2023nua,Carette:2023wpz,Castro-Ruiz:2025yvi},
superposition of quantum frames \citep{delaHamette:2021iwx}, perspective
dependence \citep{2020Quant...4..225V,Hoehn:2019fsy,Hohn_2020} and
symmetry-based formulations \citep{Loveridge:2017pcv} and large number
of relevant references therein. And the quantum reference frame relies
on a relational interpretation of quantum physics, which started from
Rovelli's seminal paper \citep{Rovelli:1995fv}. On the other hand,
through analogies between the Hilbert space and the Riemannian spaces
endowed with additional geometric structures (such as complex and
symplectic structures), numerous attempts have been made to geometrize
quantum mechanics and incorporate it into the realm of pure geometry
(see \citep{Kibble:1978tm,Provost:1980nc,Ashtekar:1997ud,Brody:1999cw,Grabowski:2005my,Bengtsson:2006rfv}
and the references therein). Among these, \citep{Pandya:2006mrv,Matassa:2020dnc}
place particular emphasis on the application of the Fubini-Study metric,
a special metric in the Complex Projective Space, in describing quantum
states, as well as the geometry underlying non-dynamical phases of
quantum states \citep{Simon:1983mh,Page:1987ca,Anandan:1990fq,Artacho:2017zey}.
Meanwhile, \citep{Mosseri:2001gbm,Levay:2003pb,Grabowski:2005my}
focus on understanding quantum entangled states from a geometric perspective.
There have been many endeavors to establish an intrinsic, background-independent,
and geometrized quantum mechanics. \citep{Minic:2003en,Minic:2003nx}
propose a quantum evolution equation akin to Einstein's equation and
provide a geometrized interpretation of probabilities. The problem
of time in covariant quantum gravity is given by e.g. \citep{Rovelli:1995fv,Gambini:2006ph},
and the conditional probability interpretations of the Hamiltonian
constraint is given by e.g. \citep{Dolby:2004ak}.

This paper takes a quantum clock as an example of the simplest reference
system to explore how the motion of object's position and the readings
of the quantum clock are incorporated into a fundamental description
of quantum mechanics. Thus, the quantum mechanics discussed in this
paper is of a specific type. When compared with the functional integral
approach \citep{Luo:2021zpi,2024arXiv240809630L}, this method based
on entangled states or relative states is equivalent, yet each has
its own advantages. The relative state approach discussed in the paper
faces difficulties in being generalized to scenarios involving quantum
spacetime reference frames \citep{Luo2014The,Luo2015Dark,Luo:2015pca,Luo:2019iby,Luo:2021zpi,Luo:2022goc,Luo:2022statistics,Luo:2022ywl,Luo:2023eqf,2023AnPhy.45869452L,2024arXiv240809630L,Luo:2025cer}
and quantum field theory, whereas the functional integral approach
is relatively straightforward in this regard. The functional method
also more readily reveals the relationship between inertial forces
and gravity that arises from general coordinate systems. However,
this canonical quantum mechanical description based on relative states
is more suitable for conceptually demonstrating its connections to
the standard Schr\"{o}dinger equation in quantum mechanics and to
the standard absolute probability interpretation of the contemporary
quantum mechanics than the functional integral approach, which is
the main motivation of the paper.

In Section II of this paper, we first examine, from a historical perspective,
how clocks, as a standard reference system, have been introduced and
evolved in physics. In Section III, we discuss how to interpret the
relative relationship between the quantum state of object's coordinates
and the quantum state of a clock from the perspective of the entangled
states. We also explore the general relative interpretation of the
entangled states and its underlying geometry: the non-trivial fiber
bundles. In Section IV, from a purely geometric standpoint, we discuss
how the quantum state of object's coordinates evolves relative to
the quantum state of the clock. In Section V, the relationship between
the derived evolution equation and the Schr\"{o}dinger equation is
discussed. In Section VI, several non-inertial effects that arise
in the reference frame of this quantum clock is discussed. We give
some comparison between our work and the past works in Section VII.
Finally, we conclude the paper with a summary. In the appendix, we
review the functional integral approach, which also deals with quantum
clocks, for the purpose of comparing it with the relative state approach.

\section{Clock Time as a Reference System}

Before going into details how a clock time measuring the so-called
``time'', let us first take a broad look at the concept of spacetime
as a reference system based on the quantum equivalence principle \citep{2024arXiv240809630L}.
In fact, we have never ``really'' measured spacetime itself, all
our measurements of the so-called spacetime are indirect, relying
on some material reference system (such as light or other material
fields serving as references). A quantum equivalence principle \citep{2024arXiv240809630L}
provides the physical foundation for establishing a quantum reference
frame theory through the measurement of the spacetime using material
reference systems. Based on this principle, we know how to use material
measuring rods and clocks to measure those universal properties (independent
of the mass and other properties of the material reference system)
that can be reinterpreted as the geometric properties of the spacetime,
rather than merely properties of the measuring rods and clocks themselves.
These universal properties of the spacetime reflected in material
reference systems can manifest not only at the classical level as
first-order moment averages independent of the material reference
system's mass (such as the mean value of coordinate) but also at the
quantum level as second-order moment quantum fluctuations of the material
reference system (such as the broadening/variance of coordinate).
Attributing these universal properties of material reference systems
to spacetime may stem from human's priori instinct or a desire for
conceptual simplicity, akin to how, in early human history, barter
trading (``trading goods for goods'') was inconvenient when comparing
the values of different items. Subsequently, the development of a
universal currency, which abstracts and measures the general value
in all transactions and serves as a general equivalent, greatly simplified
trading. Similarly, the spacetime reference frame plays the role of
a general equivalent for comparing different motions. Clock time,
as the simplest example of a spacetime reference frame, is the focus
of this paper. After clarifying the example of the clock time, the
concept of spacetime reference frames can naturally be extended in
a similar manner.

The establishment of Newton's laws of motion is based on the ability
to physically measure time using mechanical clocks. Galileo discovered
the isochronism of pendulums, which could serve as the operational
principle for mechanical clocks. Although there was no ``standard
clock'' at that time to explain to Galileo why pendulums were ``isochronous'',
he could only rely on the only ``standard'' available to him: his
pulse. By Newton's era, the concept of time had been mathematized.
Analogous to the parameterization of complex curves, Newton postulated
the existence of a mechanical device, akin to Galileo's pendulum,
that automatically ``generates'' a parameter---(Newtonian) time---for
the moving objects under study. When seemingly complex motions in
nature are parameterized by this time variable, they are greatly simplified.
To establish his laws of motion, Newton's first step was to define
a clock through his first law. Under this defined clock, the equation
of Newton's second law takes on the simplest form. Newton's laws of
motion aim to determine the forms of the spatial coordinates $X$,
$Y$, and $Z$ of an object moving in three-dimensional space when
parameterized by the Newtonian time $t$, denoted as $X(t)$, $Y(t)$,
and $Z(t)$, which are referred to as the equations of motion for
the object.

However, after people gradually gained an understanding of electromagnetism
through experiments in the 19th century, the framework of the parameter---generated
by Newton using a pendulum---came under challenge. This is because
Maxwell's equations in electromagnetism do not satisfy Galileo's principle
of relativity. When Maxwell's equations are transformed into a uniformly
moving coordinate system, their mathematical form changes. In other
words, it seemed possible to determine, through electromagnetic experiments,
whether an observer was undergoing absolute motion, which contradicted
the relativity principle that motion is only relative. So, was it
Maxwell's equations that were incorrect in moving coordinate systems,
or was it Galileo's principle of relativity that was flawed? It was
not until Einstein that the crux of the problem became clear: to ensure
that Maxwell's equations and the principle of relativity both hold
true in moving coordinate systems, Newton's notion of the global parameter
$t$ representing time had to be abandoned. In Einstein's view, time
is the value $T$ that an observer reads from his/her clock, rather
than Newton's parameter $t$. Moreover, the clock reading $T$ that
an observer obtains varies across coordinate systems moving at different
velocities, provided that the speed of light is a constant in all
moving frames. Thus, to describe the motion of an object, in addition
to the three spatial coordinates $X(\tau)$, $Y(\tau)$, and $Z(\tau)$,
there must be also Einstein's clock reading $T(\tau)$, assuming the
existence of a global parameter $\tau$. This hypothetical global
parameter $\tau$ merely serves as a common parameter for $X(\tau)$,
$Y(\tau)$, $Z(\tau)$, and $T(\tau)$, and does not necessarily have
to be related to time any more, although we often refer to it as the
``proper time'' for historical reasons. Einstein discarded Newton's
unobservable global parameter $t$ as time and instead used the physical
clock readings $T$ in coordinate systems moving at different velocities
as an intermediate equivalent for comparing different motions within
the same inertial frame: (Einsteinian) time. Consequently, the equations
of motion for objects in inertial frames moving at different velocities
transformed from a set of functions $X(t)$, $Y(t)$, and $Z(t)$
into a set of functionals $X[T(\tau)]$, $Y[T(\tau)]$, and $Z[T(\tau)]$.

Although Einstein redefined time in a stricter sense within physical
systems, Newton's global parameter $t$ still persists in the contemporary
quantum theory presented in current textbooks. In quantum mechanics,
Heisenberg was the first to abandon the concept of unobservable electron
orbits within atoms, instead focusing solely on observable quantities,
such as atomic spectral lines. In Bohr's early interpretations, since
each spectral line arises from transitions between two electron energy
states, so in Heisenberg's framework, the electron's spatial coordinates
cease to be ordinary classical numbers (c-numbers) and instead become
matrices (q-numbers) representing transitions between different energy
states. In other words, the electron's spatial coordinates transform
from a single number into a matrix. Matrix mechanics marked the first
step towards the textbook quantum mechanics we recognize today. Although
the spatial coordinates of microscopic particles are re-interpreted
in quantum mechanics, time remains Newton's global parameter. To ensure
that quantum theory satisfies Lorentz invariance and holds true in
Lorentz inertial frames, quantum field theory merely replaces Newton's
single global parameter $t$ with four global parameters $x,y,z,t$
interpreting them as Minkowski spacetime coordinates. Quantum mechanics
presupposes a classical external observer wielding Newton's parameter
$t$ generator, while quantum field theory's classical external observer
generates four global parameters. The existence of such external observers
leads to fundamental difficulties in the contemporary quantum theory.
For instance, it cannot be applied to the entire universe because,
by definition, the universe has no exterior. Moreover, quantum theory's
division of the world into an observed quantum system and an external
classical observer/instrument necessitates additional assumptions
in contemporary quantum theory to handle the measurement process between
the external classical observer and the observed quantum system, such
as the collapse of wave function in the Copenhagen interpretation.

Einstein's theory of relativity informs us that time is no longer
Newton's global external parameter $t$, but rather the clock reading
$T(\tau)$ in each inertial frame. A clock is a physical instrument
that ``generates'' a standardized motion $T(\tau)$ as an idealized
reference, such as the uniform motion of an object described by Newton's
first law or the uniform circular motion of a clock's pointer. However,
we know from quantum theory that all motions in the world are non-idealized
because they are always subject to quantum fluctuations. In other
words, there may not exist an absolutely idealized, fluctuation-free
quantum clock in the world. Nevertheless, the fact that an object
serving as a clock is always subject to quantum fluctuations may not
be of great significance, since the clock merely acts as an intermediate
equivalent or reference point to facilitate comparisons between other
motions, thereby simplifying complex motions. It is analogous to how,
when exchanging goods directly, the values of the goods can be complex
and difficult to discern. However, when a common intermediate equivalent,
``money'', is used for exchanges, the values of the goods become
immediately apparent. The fluctuations in the value of the ``money''
itself are not important, what matters is the relationship between
different goods and different motions.

To mathematically define a quantum clock as a reference system for
comparing different motions, the first question that arises is how
to quantize Einstein's clock $T(\tau)$ in such a way that it not
only adheres to quantum mechanics, experiencing quantum fluctuations,
but also satisfies the principle of relativity. Additionally, the
second question we must pose is: what is the precise meaning, at the
quantum level, of the classical-level functional equations of motion
$X[T(\tau)]$, $Y[T(\tau)]$, and $Z[T(\tau)]$?

\section{Entangled State as a relative state}

For simplicity, let us consider a one-dimensional moving object with
a spatial coordinate $X(\tau)$, whose motion with respect to the
parameter $\tau$ is governed by a Hamiltonian $H_{X}$. Additionally,
we now introduce a clock, with its pointer coordinate $T(\tau)$ evolving
with respect to $\tau$ as described by a Hamiltonian $H_{T}$. Here,
the global parameter $\tau$ shared by both $X$ and $T$ can be interpreted
as a global Newtonian parameter. Furthermore, we assume that there
is no interaction between the object $X(\tau)$ and the clock $T(\tau)$.
They evolve independently with respect to the parameter $\tau$. Thus,
the Hamiltonian of the entire system is given by $H=H_{X}+H_{T}$.
The quantum state space of the entire system is a direct product space
\begin{equation}
\mathscr{H}_{X}\otimes\mathscr{H}_{T}
\end{equation}
where $\mathscr{H}_{X}$ is the quantum state space spanned by the
object's coordinate $X(\tau)$ and $\mathscr{H}_{T}$ is the quantum
state space spanned by the clock's pointer coordinate $T(\tau)$.
However, the quantum state of the entire system, as an eigenstate
of the total Hamiltonian, is not necessarily a simple direct product
state of the object's quantum state $|X(\tau)\rangle$ and the clock's
quantum state $|T(\tau)\rangle$. Generally, it is an entangled state,
with a dimensionality $N$ equal to that of the smaller Hilbert subspace
\begin{equation}
|X,T\rangle=\sum_{\tau}^{N}C_{\tau}|X(\tau)\rangle\otimes|T(\tau)\rangle\label{eq:entangled state}
\end{equation}

All measurement processes involve a comparison between the state of
the system under-study and the state of the measuring instrument.
Prior to conducting a measurement on the system, an initialization
step known as instrument calibration (or scale mark) is necessary.
This step involves adjusting the states of the object's coordinate
$X$ and the clock $T$ to establish a one-to-one correspondence between
the instrument (the clock $T$) and the system under-study (the object
position $X$), thereby enabling the measurement of the system's state
by reading the instrument's state. The ``one-to-one correspondence''
is equivalent to being able to extract the maximum information from
the under-study system. The superposition coefficients $C_{\tau}$
of the entangled state (\ref{eq:entangled state}) are determined
by the state preparation during calibration. For instance, the system
can be adjusted so that at $\tau=0$, the clock pointer is initialized
in the state $T(0)=0$, while the object's coordinate is initialized
at the position $X(0)=X_{0}$. This process is equivalent to subjecting
them to an instantaneous interaction, after which they cease to interact
and instead evolve independently according to their respective Hamiltonians
with respect to $\tau$. During subsequent evolution, whenever the
clock is in a certain evolved state $|T(\tau)\rangle$, the object
is also necessarily in a corresponding evolved state $|X(\tau)\rangle$.
Thus, the state of the entire system is at $|X(\tau)\rangle\otimes|T(\tau)\rangle$,
related to an amplitude of $C_{\tau}$ in front and a probability
of $|C_{\tau}|^{2}$. The amplitude or probability that the object
is in a corresponding state when the clock is in a certain state is
prepared during instrument calibration. Moreover, quantum principles
permit the existence of superposition of these one-to-one states,
where all possible one-to-one states of the clock and object position
are superposed with the corresponding amplitudes $C_{\tau}$, resulting
in the entangled state (\ref{eq:entangled state}). From a mathematical
perspective, this entangled state establishes a one-to-one (probabilistic)
mapping $|X,T\rangle:|T(\tau)\rangle\rightarrow|X(\tau)\rangle$ between
the object's position states $|X(\tau)\rangle$ under possible clock
states $|T(\tau)\rangle$. This mathematical mapping is essentially
the counterpart of the functional mapping $X[T(\tau)]:T(\tau)\rightarrow X(\tau)$,
so in this sense, the entangled state can be regarded as a quantum
version of the functional equation of motion $X[T(\tau)]$.

Unlike deterministic events described by functional equations of motion
$X[T(\tau)]$, this entangled state depicts a probabilistic event.
When the clock is in the state $|T(\tau)\rangle$, the object's position
is in the state $|X(\tau)\rangle$. The normalized joint probability
of these two events occurring simultaneously is $|C_{\tau}|^{2}$
(w.r.t. an external observer). Unlike a simple direct product state
$|T\rangle\otimes|X\rangle$, the entangled state $|X,T\rangle$ is
inseparable, meaning that the joint probability $|C_{\tau}|^{2}$
does not equal the simple product of the normalized probability $|A_{\tau}|^{2}$
of measuring the clock alone in the state $|T(\tau)\rangle$ and the
normalized probability $|B_{\tau}|^{2}$ of measuring the object's
position alone in the state $|X(\tau)\rangle$, i.e. $|C_{\tau}|^{2}\neq|A_{\tau}|^{2}|B_{\tau}|^{2}$.
Here, $|T\rangle=\sum_{\tau}A_{\tau}|T(\tau)\rangle$ and $|X\rangle=\sum_{\tau}B_{\tau}|X(\tau)\rangle$
represent certain quantum states that each subsystem alone would be
in, expanded in terms of their respective basis vectors in the basis
of the entangled state. Therefore, the probability of the object's
position being in the state $|X(\tau)\rangle$ given that the clock
is definitely in the state $|T\rangle$ is a conditional probability
(or relative probability), which can be derived using the conditional
probability formula
\begin{equation}
\textrm{P}\left(X(\tau)|T\right)=\frac{\textrm{P}\left(X(\tau)\cup T(\tau)\right)}{\textrm{P}\left(T(\tau)\right)}=\frac{\textrm{P}\left(X(\tau)\cup T(\tau)\right)}{\textrm{P}\left(T(\tau)|T\right)}=\frac{|C_{\tau}|^{2}}{|A_{\tau}|^{2}}.\label{eq:conditional pro}
\end{equation}
where $\textrm{P}\left(T(\tau)\right)=\textrm{P}\left(T(\tau)|T\right)=|A_{\tau}|^{2}$
represents the probability (w.r.t. an external observer) that the
reference state $|T(\tau)\rangle$ is found in the subsystem state
$|T\rangle$ of the entangled state, while $\textrm{P}\left(X(\tau)\cup T(\tau)\right)=|C_{\tau}|^{2}$
denotes the joint probability (w.r.t. an external observer) that the
entangled state is simultaneously in the states $|X(\tau)\rangle$
and $|T(\tau)\rangle$. However, their ratio, the relative probability
is then independent to the external observer.

We refer to the quantum state $|T\rangle$ of the quantum clock or
measuring instrument as the conditional state (or reference state).
It is input by observer and depends on the specific conditions under
which the observer wishes to obtain the relative probabilities of
the entangled state $|X,T\rangle$ as output. This constitutes the
``relative probability'' interpretation of the entangled states.
It is a distinct interpretation from the ``absolute probability''
interpretation in the standard quantum mechanics, as it additionally
takes into account the conditions of the state occupied by the instrument's
subspace. Within the framework, to extract physically meaningful measurement
results from the entangled state that describes the relative relationship
between the under-study quantum system and the quantum instrument,
one must rely on the relative probability interpretation.

We observe that the entangled state $|X,T\rangle$ cannot be expressed
as a direct product state, i.e. $|X,T\rangle\neq|X\rangle\otimes|T\rangle$.
The relative probability of the object's position being in the state
$|X(\tau)\rangle$ given that the clock is in the state $|T\rangle$
does not equal the probability obtained from a separate measurement
of the object's position, i.e. $\textrm{P}\left(X(\tau)|T\right)=\frac{|C_{\tau}|^{2}}{|A_{\tau}|^{2}}\neq|B_{\tau}|^{2}$.
Therefore, the inseparable entangled state $|X,T\rangle$ can only
be interpreted probabilistically based on the relative relationship
between the two states. The state of the object's position $|X(\tau)\rangle$
only holds physical meaning relative to the state of the clock $|T(\tau)\rangle$.
Discussing the state of the object's position alone or the state of
the clock alone lacks absolute meaning, as their exhibited quantum
probabilistic behaviors are entirely different.

The inseparability of the entangled state (\ref{eq:entangled state}),
or its inability to be expressed as a global direct product state,
mathematically means that the entangled state $|X,T\rangle$ cannot
be represented by a single global clock state that encompasses all
possible position states. Instead, it can only locally map out a position
state $|X(\tau)\rangle$ on a local clock state $|T(\tau)\rangle$,
and then superpose (or ``glue together'') the states of all local
bases to obtain the overall quantum entangled state. In other words,
the entangled state $|X,T\rangle=\sum_{\tau}C_{\tau}|X(\tau)\rangle\otimes|T(\tau)\rangle$
forms a non-trivial fiber bundle with a fiber $|X(\tau)\rangle$ growing
over the local base space $|T(\tau)\rangle$, and a projection $T^{-1}$
\begin{equation}
T^{-1}:|X,T\rangle\rightarrow|X(\tau)\rangle\label{eq:fiber bundle projection}
\end{equation}
from the fiber bundle $|X,T\rangle$ to a section $|X(\tau)\rangle$
of the local fiber, gives rise to the relative amplitude (see below).
The fiber bundle $|X,T\rangle$ is non-trivial means that it is not
equivalent to a globally direct product state $|X\rangle\otimes|T\rangle$,
i.e. the entangled state $|X,T\rangle$ corresponds to a non-trivial
fiber bundle. Compared with that, in the standard quantum mechanics,
the wave function $|X(t)\rangle$ is trivially a global section of
the fiber over a global Newtonian time $t$ as the base space, meaning
that it is a trivial fiber bundle.

Only when the probability of the object's position state $\textrm{P}\left(X(\tau)|T\right)$,
equals the probability obtained from separate measurements of the
object's position and the clock, i.e., when the state is separable
i.e. $|C_{\tau}|^{2}=|A_{\tau}|^{2}|B_{\tau}|^{2}$, can we use a
single global time to cover all possible position states. In this
case, the entangled state $|X,T\rangle$ reverts to the separable
direct product state $|X\rangle\otimes|T\rangle$ or a trivial fiber
bundle, allowing the entire background state of the clock to be separated
out and reduced to the standard textbook quantum state $|X\rangle$,
which describes the probability $|B_{\tau}|^{2}$ of measuring the
object's position $X$ at a global time $\tau$ ($|B_{\tau}|^{2}$
has physical meaning w.r.t. $\tau$). Therefore, we require a relativistic
quantum mechanical interpretation based on the ``relative relationship
between the two states $|X(\tau)\rangle$ and $|T(\tau)\rangle$''
described by the local entangled states, to generalize the Copenhagen
interpretation of quantum mechanics for ``absolute quantum states
$|X(\tau)\rangle$'' in textbooks. This is analogous to generalizing
global Cartesian geometry to curved Riemannian geometry based on tangent
bundles over local base spaces. Establishing a relativistic quantum
theory and an intrinsic interpretation that measures one entangled
subspace (the quantum system under-study) relative to another subspace
(the quantum measuring instrument) within a larger quantum state space
is particularly urgent in explaining a quantum theory of gravity.

An important consequence of the distinction between the non-trivial
fiber bundle described by the entangled states and the global trivial
fiber bundle depicted in conventional quantum mechanics lies in the
fact that, due to the absence of a global base space in the non-trivial
fiber bundle, the local basis vectors on the base space $T$ are not
necessarily orthogonal (i.e. nonorthogonal). Instead, they define
a non-trivial metric on the base space $T$ given by the inner product
$\langle T(\tau)|T(\tau^{\prime})\rangle=s^{\tau\tau^{\prime}}$,
which is not necessarily gauge equivalent to the Kronecker's delta
$\delta_{\tau\tau^{\prime}}$ of the flat case
\begin{equation}
\langle T(\tau)|T(\tau^{\prime})\rangle=\langle T(\tau)|T\rangle\langle T|T(\tau^{\prime})\rangle=A^{\tau*}A^{\tau^{\prime}}=\frac{A^{\tau*}}{A_{\tau^{\prime}}}=A^{\tau}A^{\tau^{\prime}*}=\frac{A^{\tau}}{A_{\tau^{\prime}}^{*}}=s^{\tau\tau^{\prime}}=\left(s_{\tau\tau^{\prime}}\right)^{-1}\neq\delta_{\tau\tau^{\prime}}\label{eq:<TT>}
\end{equation}
In other words, the base space of a non-trivial fiber bundle can be
seem ``curved'', unlike the ``flat'' basis vectors of the global
trivial fiber bundle in conventional quantum mechanics. The orthonormality
or invariant inner product must now be expressed in terms of the basis
vectors $|T\rangle$ and their dual basis vectors $\langle\bar{T}|$
\begin{equation}
\langle\bar{T}(\tau)|T(\tau^{\prime})\rangle=A_{\tau}^{*}A^{\tau^{\prime}}=\langle T(\tau)|\bar{T}(\tau^{\prime})\rangle=A_{\tau}A^{\tau^{\prime}*}=\delta_{\tau}^{\tau^{\prime}}\label{eq:TbarT}
\end{equation}
As long as the expansion coefficients $A_{\tau}$ are normalized,
it follows that $\sum_{\tau}A_{\tau}^{*}A_{\tau}=\sum_{\tau}s_{\tau\tau}=1$.

The second method for calculating the relative amplitude is by using
the projection of the fiber bundle (\ref{eq:fiber bundle projection}):
the relative amplitude is obtained by performing a ``conditional
projection'' of the vector $|X,T\rangle$ in the non-trivial fiber
bundle onto the section $|X(\tau)\rangle$ of the fiber, where the
reference state $|T\rangle$ as a projector can be expanded by using
the basis vectors $|T(\tau)\rangle$ in the entangled state
\begin{align}
\langle T| & =\sum_{\tau}A_{\tau}^{*}\langle T(\tau)|=\sum_{\tau,\tau^{\prime}}A_{\tau}^{*}s^{\tau\tau^{\prime}}\langle\bar{T}(\tau^{\prime})|=\sum_{\tau,\tau^{\prime}}A_{\tau}^{*}\left(A^{\tau^{\prime}}A^{\tau*}\right)\langle\bar{T}(\tau^{\prime})|\nonumber \\
 & =\sum_{\tau,\tau^{\prime}}\delta_{\tau}^{\tau^{\prime}}A^{\tau*}\langle\bar{T}(\tau^{\prime})|=\sum_{\tau}A^{\tau^{\prime}*}\langle\bar{T}(\tau^{\prime})|=\sum_{\tau^{\prime}}\frac{1}{A_{\tau^{\prime}}}\langle\bar{T}(\tau^{\prime})|
\end{align}
By using the conditional projection by $\langle T|$, we have a projected
entangled state 
\begin{equation}
\langle T|X,T\rangle=\sum_{\tau}C_{\tau}|X(\tau)\rangle\langle T|T(\tau)\rangle=\sum_{\tau,\tau^{\prime}}C_{\tau}A_{\tau^{\prime}}^{*}\langle T(\tau^{\prime})|T(\tau)\rangle|X(\tau)\rangle=\sum_{\tau,\tau^{\prime}}C_{\tau}A_{\tau^{\prime}}^{*}s^{\tau\tau^{\prime}}|X(\tau)\rangle=\sum_{\tau}\frac{C_{\tau}}{A_{\tau}}|X(\tau)\rangle\label{eq:projected entangled state}
\end{equation}
in which the coefficient $\frac{C_{\tau}}{A_{\tau}}$ describes the
relative amplitude (or conditional amplitude) of measuring the object's
position in the state $|X(\tau)\rangle$ given that the clock is in
the state $|T\rangle$, reflecting the relative amplitude between
the subsystems, then yielding the relative probability $\frac{|C_{\tau}|^{2}}{|A_{\tau}|^{2}}$.

Now, the magnitude (or the length) of the component $C_{\tau}$ of
the state vector in the space $\mathscr{H}_{X}\otimes\mathscr{H}_{T}$
no longer holds absolute meaning (relative to the absolute normalization
factor $\sum_{\tau}|C_{\tau}|^{2}$ and external parameter), instead,
it only has relative meaning with respect to the magnitude $|A_{\tau}|^{2}$
of the component $A_{\tau}$ in the subspace $\mathscr{H}_{T}$. More
precisely, what we refer to in this paper as describing the ``relative
relationship'' between two mutually entangled systems essentially
involves characterizing the ``relative component'' of the entangled
state vector in one subspace relative to the other subspace. This
``relative component'' is precisely the relative amplitude or conditional
amplitude.

The third method for calculating the relative amplitude or relative
probability involves noting that this relative probability is also
equivalent to taking the partial trace of the density matrix $\rho_{X,T}\equiv|X,T\rangle\langle X,T|$
of the entangled state with respect to the reference state $|T\rangle$
\begin{equation}
\textrm{Tr}_{T}\left(\rho_{X,T}\right)=\langle T|X,T\rangle\langle X,T|T\rangle=\sum_{\tau,\tau^{\prime}}\frac{C_{\tau}}{A_{\tau}}\frac{C_{\tau^{\prime}}^{*}}{A_{\tau^{\prime}}^{*}}|X(\tau)\rangle\langle X(\tau^{\prime})|=\sum_{\tau}\left|\frac{C_{\tau}}{A_{\tau}}\right|^{2}|X(\tau)\rangle\langle X(\tau)|=\sum_{\tau,\tau^{\prime}}C_{\tau}C_{\tau^{\prime}}^{*}s^{\tau\tau^{\prime}}|X(\tau)\rangle\langle X(\tau^{\prime})|
\end{equation}
Here, taking the partial trace of the entangled state's density matrix
is not simply summing the diagonal elements within a subspace, but
rather contracting by the metric $s^{\tau\tau^{\prime}}$of the subspace
$T$, i.e. $\textrm{Tr}_{T}\left(\mathbf{O}\right)=\mathbf{O}_{\mu\nu}s^{\mu\nu}\neq\mathbf{O}_{\mu\nu}\delta_{\mu\nu}$.
This is analogous to how, in a curved Riemannian space, taking the
trace of a matrix is no longer a straightforward summation of all
diagonal elements, instead, it requires contraction using the curved
metric. This represents a significant difference between this approach
and the conventional (global) quantum mechanics. 

Because the amplitude of the entangled state, $C_{\tau}$, is determined
by the calibration and/or preparation of the instrument, rather than
by the fundamental equations of theory, and remains fixed during the
experimental process, so the entangled state merely serves as a ``cross-reference
table'' between the instrument readings and the state of the system
under-study prepared during the calibration of the system under-study
and the quantum instrument. The relative amplitude obtained through
metric projection onto the conditional state, as well as its evolution
(relative to the instrument subsystem), are entirely determined by
the metric of one subsystem $X$ and its evolution relative to another
subsystem $T$ of the entangled state. The relative probability given
by the relative amplitude is physically measurable. The evolution
of the metric of the subspace $X$ relative to $T$ is coming from
the Ricci-flat K\"{a}hler-Einstein equation (see the next chapter).
This equation links the intrinsic curvature of the metric of the non-flat
subspace $\mathscr{H}_{X}$ embedded in the flat quantum state space
$\mathscr{H}_{X}\otimes\mathscr{H}_{T}$ with the extrinsic curvature
relative to the evolution of the another subspace $\mathscr{H}_{T}$.
It serves as the equation of motion for the metric of one subspace
$\mathscr{H}_{X}$ relative to the state vectors of the other subspace
$\mathscr{H}_{T}$. Similarly, we also have the evolution equation
for the metric of the subspace $\mathscr{H}_{T}$ relative to $\mathscr{H}_{X}$,
which describes its relative evolution with respect to the state vectors
of the subspace $\mathscr{H}_{X}$. The system under-study and the
measuring instrument now are completely on an equal footing. More
discussion on the relationship between evolution equation and its
degradation to the Schr\"{o}dinger equation will be provided in the
following paper. Thus, we observe that the evolution of the relative
amplitude is not generated by the evolution of the entangled state
(which remain fixed after the calibration), but rather by the relative
evolution of the subspace metrics or basis vectors of the Hilbert
subspaces relative to another subspace. This differs from standard
quantum mechanics, where the initially prepared quantum state undergoes
unitary evolution (unitary rotation) relative to external parameters.
Here, the entangled state vector does not evolve relative to external
parameters, instead, relative evolution occurs between Hilbert subspaces,
causing the components of the entangled state $C_{\tau}$ projected
onto the subspace $\frac{C_{\tau}}{A_{\tau}}$ (or $\frac{C_{\tau}}{B_{\tau}}$)
to evolve relative to another subsystem. Adopting the form of the
evolution equation for the intrinsic representation of quantum states
allows us to treat the system under-study (object's position $X$)
and the measuring instrument (physical clock $T$) on an equal footing
and symmetrically, in other words, we can also treat the physical
clock as the system under-study and the object's position as the measuring
instrument.

The derivative of the basis vector $|T(\tau)\rangle$ of the base
space with respect to $\tau$, namely $\langle T(\tau)|\frac{d}{d\tau}|T(\tau)\rangle$,
yields a non-trivial connection. This non-trivial connection on the
curved base space leads to the emergence of a non-integrable Berry
phase $\Delta\varTheta$ when the state vector undergoes parallel
transport across the base space
\begin{equation}
\Delta\varTheta=\int_{0}^{\hat{\tau}}d\tau\langle T(\tau)|\frac{d}{d\tau}|T(\tau)\rangle=\oint_{c}d\mathbf{x}(\tau)\cdot\langle T[\mathbf{x}(\tau)]|\mathbf{\nabla}_{\mathbf{x}}|T[\mathbf{x}(\tau)]\rangle=\oint_{c}d\mathbf{x}(\tau)\cdot\mathbf{K}[\mathbf{x}(\tau)]
\end{equation}
The non-vanishing of the phase change along a closed loop means that
there is no longer a global direct product of state vector, the state
formed by these two subsystems must be an entangled state. (To facilitate
the observation of the relationship with the standard Berry phase,
we can define $\mathbf{x}(\tau)$ as a higher-dimensional parameter
space corresponds to $\tau$. The connection of the state vector as
the parameters change is $\langle T[\mathbf{x}(\tau)]|\mathbf{\nabla}_{\mathbf{x}}|T[\mathbf{x}(\tau)]\rangle\equiv\mathbf{K}[\mathbf{x}(\tau)]$.
Starting from $\mathbf{x}(0)$ and traversing a closed loop $c$ in
the higher-dimensional parameter space, we return to $\mathbf{x}(\hat{\tau})=\mathbf{x}(0)$.
Here, $\mathbf{B}=\nabla_{\mathbf{x}}\times\mathbf{K}[\mathbf{x}(\tau)]$
represents the non-zero Berry curvature in the $\mathscr{H}_{T}$
subspace). Therefore, we observe that, in general, the inherent nature
of non-trivial fiber bundles in the subsystem of an entangled state
ensures the existence of a non-trivial Berry phase in the subspace
and the subspace must be curved, in other words, the first Chern class
or Chern number is non-trivial in the base space of the non-trivial
fiber bundles.

The entangled state between the under-study system and that of a quantum
instrument, along with its interpretation in terms of the relative
probabilities, provides the conceptual foundation for a quantum theory
formulated in a general spacetime coordinate system. Entangled states
are well-defined within the framework of quantum theory, yet they
exhibit many properties that appear strange from the perspective of
classical physics. In the early years of the quantum mechanics, it
was assumed that the scale of the wave function was limited to atomic
scale, and thus there was no need to worry about the issue of simultaneity
arising from the instantaneous collapse of the wave function in the
Copenhagen interpretation. However, Einstein's EPR thought experiment
pointed out that the scale of the wave function could, in fact, be
much larger---so large that the wave functions are completely macroscopically
orthogonal in space. This leads to the perplexing problem of ``superluminal''
effects associated with the instantaneous collapse of the wave function.
Here, we merely offer an alternative perspective on the entangled
state, viewing them as a type of ``relative state'' that describes
the ``relationship'' between two states. This approach makes the
picture of entangled state less incomprehensible. Take, for example,
the spin version of the EPR thought experiment. A pair of spin-polarized
photons or electrons, after undergoing a brief interaction---such
as when light passes through a nonlinear medium, emitting two oppositely
moving polarized photons A and B, leading to an entangled state, often
called, the EPR state
\begin{equation}
|\textrm{EPR}\rangle=\frac{1}{\sqrt{2}}\left(|\mathbf{n}\rangle_{A}\otimes|-\mathbf{n}\rangle_{B}\pm|-\mathbf{n}\rangle_{A}\otimes|\mathbf{n}\rangle_{B}\right)
\end{equation}
where $\mathbf{n}$ and $-\mathbf{n}$ represent a pair of opposite
spin directions for the particles in space. We avoid the conventional
EPR state notation $|\textrm{EPR}\rangle=\frac{1}{\sqrt{2}}\left(|\uparrow\rangle_{A}\otimes|\downarrow\rangle_{B}\pm|\downarrow\rangle_{A}\otimes|\uparrow\rangle_{B}\right)$
because it misleadingly suggests that the local states are definitively
up $\uparrow$ or down $\downarrow$. In reality, when we locally
measure the spin of particle A, we find that its spin $\mathbf{n}$
is completely random. Similarly, measuring the spin of particle B
alone also yields a completely random result. Although the spin directions
of A and B are individually random, they are always opposite to each
other, a fact that can be verified through a joint measurement. This
indicates that the individual spin quantum states of particles A or
B do not have absolute meaning. What is physically meaningful is the
relationship between the spins of the two particles. The EPR state
describes only this relationship, not the local properties of the
particles. Just as it is meaningless to discuss the absolute position
of an object without a reference frame or coordinate system, what
is physically meaningful is the object's position relative to the
origin of a coordinate system. Therefore, our theory aims to describe
not the spin states of individual local particles but rather the relationship
between the spin states of the two particles. From this perspective,
there is no need to worry about whether the collapse of the entangled
state involves mysterious ``superluminal'' effects, because what
``collapses'' is a certain relationship, not local quantum properties.
(In this sense, the interpretation of quantum collapse probability
here is somewhat closer to the Bayesian notion of ``a measurement
of the degree of belief'' as the ``relationship'' is more akin
to an observer's constructed belief about the objective reality. However,
a ``relationship'' is of course not entirely subjective, the relationship
between the two is an objective existence.) From the standpoint of
the standard quantum state space of the overall entangled state, the
``interrelationship'' between the spin states of the two particles
does indeed appear to collapse instantaneously to an external observer,
just as the standard Copenhagen interpretation suggests. This does
not violate relativity because there is no absolute external observer
who can observe the ``collapse'' of the relative state from the
outside. When the entangled state collapses to certain its component,
the relationship between the two particles becomes deterministic for
the ``external observer'' but the specific local individual spin
of A or B (for the external observer) remain undetermined. In other
words, this interrelationship does not inform observers at A and B
about the absolute spin directions of their respective particles.
The spin quantum states of A and B still require specific local measurements
by observers at each location. For example, after A measures its spin
state, A communicates this locally measured result to observer B via
a signal that travels at most at the speed of light. Given A's measurement
result, observer B can then calculate the relative probability and
expectation value for the direction of B's measured spin state. Otherwise,
B's local observer perceives the spin direction as completely random
relative to B's local measurement apparatus. In other words, we can
only measure the relative relationship between A and B, while the
absolute states of A and B locally are completely random. This is
because if you project the spin state prepared locally by a certain
instrument, say $|\mathbf{m}\rangle$ (e.g., via a Stern-Gerlach gradient
magnetic field prepared in a specific direction), the local spin state
$|\mathbf{n}\rangle$ can form an arbitrary angle with the prepared
spin state $|\mathbf{m}\rangle$, resulting in an arbitrary inner
product and relative phase. To measure the relative phase (or corresponding
relative amplitude) between A and B, the two particles must be retransmitted
at no more than the speed of light to the same point in space for
an interference experiment. This interference experiment is equivalent
to projecting the entangled state using the state of one of the subsystems
as a reference state, yielding the relative amplitude or relative
phase between them.

The relative or conditional probabilities derived from an entangled
state reflect the fact that our information about the entangled system
is incomplete in the sense that when one subsystem of the entangled
state learns, via a classical channel, the measurement result of the
other subsystem, it can utilize this additional information to update
the prediction or the (posterior) probability about this local subsystem,
as is claimed in the Bayesian theorem for an information incomplete
system, such as the game of bridge where each player can only see
a portion of the entire deck of cards.

In this EPR experiment, we can consider that the particle at location
A serves as the measuring instrument or reference frame, while the
other entangled particle B acts as the under-study system. The physically
meaningful spin direction of the particle B is only relative to the
spin direction of the measuring instrument, the particle A. Without
knowledge of the state of the measuring instrument (the particle A),
we cannot discuss the spin state of the particle B, and vice versa.
An imperfect classical analogy would be the voltage difference between
two widely separated electrodes. Suppose the voltage difference between
them is fixed at 10V. This voltage difference can only have relative
physical meaning. However, locally, the voltage measured at each electrode
is entirely random, depending on the zero-point potential selected
by the local observer's measuring instrument. When you measure your
local potential and find it to be, say, a completely random value
$a$V, and then communicate this value to another observer via a signal
traveling at a finite speed, the potential of the other observer then
``instantaneous'' becomes $(a+10)$V. This is because the voltage
has only relative meaning, but in reality, there is no physical electric
potential change experienced by the observer. Our quantum theory must
also be entirely based on describing the relationships between different
states, rather than on external absolute states. It should be founded
on describing the relationship between the quantum states of the quantum
system under-study and the quantum instrument, because, fundamentally,
the measuring instrument is also part of the quantum world. It can
exist in quantum superposition states and satisfy the quantum uncertainty
principle, while the standard quantum theory assumes that the measuring
instrument can only be in classical states and yield classical measurement
results. Only when the measuring instrument can be approximately treated
as a classical instrument, with its quantum effects ignored, do we
revert to the textbook quantum theory that describes the quantum state
of a single system being measured. Below, we will use a quantum clock
as an simplest example to illustrate how to develop a dynamics of
quantum theory upon describing the relationships between quantum states.
We generalize this simple example to more general quantum spacetime
reference frames to establish a quantum theory on more general spacetime
coordinates and gravity \citep{Luo2014The,Luo2015Dark,Luo:2015pca,Luo:2019iby,Luo:2021zpi,Luo:2022goc,Luo:2022statistics,Luo:2022ywl,Luo:2023eqf,2023AnPhy.45869452L,2024arXiv240809630L,Luo:2025cer}.

\section{Relative Evolution of States and K\"{a}hler-Einstein Equation}

We will base our more detailed discussion on the relative evolution
and ``relative relationship'' between the quantum state of the under-study
system $|X_{i}\rangle\in\mathscr{H}_{X}$ and the quantum state of
the measuring instrument or clock $|T_{a}\rangle\in\mathscr{H}_{T}$
on the fiber bundles or the geometry of the quantum state space $\mathscr{H}_{X}\otimes\mathscr{H}_{T}$
formed by the entangled state and its subspaces. This represents an
intrinsic geometric perspective that treats the under-study subsystem
and the quantum measuring instrument in a more symmetric manner.

To better highlight the similarities between the Hilbert space and
the tensorial representation of Riemannian geometry, we will adopt
a discrete basis vector formulation in this chapter, replacing the
continuous basis vector version used in the previous chapter. Specifically,
we will use $|X_{i}\rangle$ to replace $|X(\tau)\rangle$. Similarly,
$|T_{a}\rangle$ represents the discrete version of $|T(\tau)\rangle$.

We start with the calibration process between the under-study system
and the quantum instrument and prepares an entangled state
\begin{equation}
|X,T\rangle=\sum_{i}^{N}C_{i}|X_{i}\rangle|T_{i}\rangle,\quad\left(C_{i}\in\mathbb{C}\right)\label{eq:entangle state-2}
\end{equation}
This quantum state is a vector in the linear and flat quantum state
space $\mathscr{H}_{X}\otimes\mathscr{H}_{T}$. To ensure a one-to-one
correspondence between the quantum states of the two subspaces, the
dimension of the total quantum state space is taken to be the dimension
$N$ of the lower-dimensional subspace among the two. When this entangled
state is collapsed in a local direct product basis $|X_{i}\rangle|T_{i}\rangle$,
it gives a joint amplitude $C_{i}$ (it is generally a complex number)
that satisfy the normalization condition $\sum_{i}^{N}|C_{i}|^{2}=1$.

It is worth noting that the prepared entangled state (\ref{eq:entangle state-2})
is not a standard Schmidt decomposed entangled state. Here, $C_{i}$
is not Schmidt coefficient, and the local subspace basis vectors,
which are selected from the calibration process, are not necessarily
orthonormal. In contrast, the Schmidt decomposition requires selecting
an appropriate local coordinate system $|X_{i}^{\prime}\rangle|T_{i}^{\prime}\rangle$
in the total quantum state space such that the expansion coefficients
$C_{i}^{\prime}$ of the entangled state in this coordinate system
are real-valued, and the local basis vectors are orthonormal.

Since a single subspace, either $\mathscr{H}_{X}$ or $\mathscr{H}_{T}$
may no longer be locally flat, we can always consider how the basis
vectors $|X_{i}\rangle$ of the local $\mathscr{H}_{X}$ subspace
vary with respect to the local quantum states in the local $\mathscr{H}_{T}$
subspace, whose state is expanded as
\begin{equation}
|T\rangle=\sum_{a}^{N}t^{a}|T_{a}\rangle,\quad\left(t^{a}\in\mathbb{\mathbb{C}}\right)\label{eq:T state}
\end{equation}
where $|T_{a}\rangle$ represents the (non-orthogonal) basis vectors
in the local $\mathscr{H}_{T}$ subspace, and $t^{a}$ is the corresponding
normalized amplitude. What we now aim to examine is the evolution
of the basis vectors $|X_{i}\rangle$ in the local $\mathscr{H}_{X}$
subspace as they vary with the complex coordinates $t^{a}$ on the
entangled local $\mathscr{H}_{T}$ subspace.

This entangled state constitutes a non-trivial fiber bundle structure
formed by the bonding of local subsystems. Although the total quantum
state space $\mathscr{H}_{X}\otimes\mathscr{H}_{T}$ is a flat linear
space, its local subspace $\mathscr{H}_{X}$ is not flat. We consider
the portion of the state space where $\mathscr{H}_{X}$ and $\mathscr{H}_{T}$
are entangled in a one-to-one correspondence, and thus its dimensionality
is also $N$. The local metric can be defined through the inner product
or overlap of the basis vectors $|X_{i}\rangle$
\begin{equation}
h_{\bar{i}j}=\langle X_{i}|X_{j}\rangle\neq\delta_{ij},\quad h_{\bar{j}i}=\langle X_{j}|X_{i}\rangle=\langle X_{i}|X_{j}\rangle^{*}=h_{\bar{i}j}^{*}=h_{i\bar{j}}
\end{equation}
The bar over the lower index $i$ indicates that the bra vector corresponding
to this lower index in the inner product has undergone a complex conjugation
(since the physical metric is Hermitian, $h_{\bar{i}j}=h_{\bar{i}j}^{*}=h_{j\bar{i}}$,
and the interchange of indices for this complex metric is symmetric).
Similarly, the bar over the upper index $j$ denotes that the ket
vector corresponding to the upper index in the inner product is a
complex conjugate of the ket vector corresponding to the lower index
\begin{equation}
h^{i\bar{j}}=\langle X^{i}|X^{j}\rangle,\quad h^{j\bar{i}}=\langle X^{j}|X^{i}\rangle=\langle X^{i}|X^{j}\rangle^{*}=h^{i\bar{j}*}=h^{\bar{i}j}
\end{equation}
Due to the Hermiticity of the metric, the metric components vanish
when both indices are unbarred or both indices are barred. The basis
vectors that are locally dual to the local basis vectors $|X_{i}\rangle$
are now denoted as $\langle X^{j}|$, so in this notation, the bar
over the bra vector $\bar{\langle X|}$ (dual to $|X\rangle$), as
previous eq.(\ref{eq:TbarT}), can be omitted in this convention.
Throughout the following paper, the upper index and lower index dual
to each other, as the notations we often used in Riemannian geometry,
they satisfy the orthogonality and normalization conditions
\begin{equation}
\langle X^{j}|X_{i}\rangle=\delta_{i}^{j},\quad\langle X_{j}|X^{i}\rangle=\delta_{\bar{j}}^{\bar{i}}\equiv\delta_{j}^{i}\label{eq:X orthogonal}
\end{equation}
Now, $|X_{i}\rangle$ represents the basis vectors of the entangled
state within the local subspace of the under-study system. Therefore,
the quantum state $|X\rangle$ can also be expanded using this set
of local basis vectors
\begin{equation}
|X\rangle=\sum_{i}^{N}x^{i}|X_{i}\rangle,\quad\left(x^{i}\in\mathbb{\mathbb{C}}\right)
\end{equation}
in which $x^{i}$ represents the normalized complex coordinates.

Similarly, the metric of the subspace $\mathscr{H}_{T}$ can be given
by the non-orthogonal basis vector
\begin{equation}
s_{\bar{a}b}=\langle T_{a}|T_{b}\rangle\neq\delta_{ab},\quad s_{\bar{b}a}=\langle T_{b}|T_{a}\rangle=\langle T_{a}|T_{b}\rangle^{*}=s_{\bar{a}b}^{*}=s_{a\bar{b}}
\end{equation}
and 
\begin{equation}
s^{a\bar{b}}=\langle T^{a}|T^{b}\rangle,\quad s^{b\bar{a}}=\langle T^{b}|T^{a}\rangle=\langle T^{a}|T^{b}\rangle^{*}=s^{a\bar{b}*}=s^{\bar{a}b}
\end{equation}
$\langle T^{b}|$ is locally dual to $|T_{a}\rangle$ and $\langle T^{b}|$
is the discrete version of $\langle\bar{T}(\tau)|$ in eq.(\ref{eq:TbarT}). 

To derive the correct evolution equation, it is necessary to assume
that the metric signature difference between $|X\rangle$ and $|T\rangle$
subspaces is opposite. In other words, we stipulate that the $|T\rangle$
subspace is timelike, meaning that it has a negative inner product
\begin{equation}
\langle T^{b}|T_{a}\rangle=-\delta_{a}^{b},\quad\langle T_{b}|T^{a}\rangle=-\delta_{\bar{b}}^{\bar{a}}\label{eq:T orthogonal}
\end{equation}
Meanwhile, $|X\rangle$ is spacelike and satisfies (\ref{eq:X orthogonal}).
The orthogonality condition (\ref{eq:T orthogonal}) does not contradict
the previously defined orthogonality in eq.(\ref{eq:TbarT}) because
we could just as well reverse the definitions entirely, assigning
$|X\rangle$ a negative inner product and $|T\rangle$ a positive
inner product instead. The roles of $|X\rangle$ as the under-study
system and $|T\rangle$ as the measuring apparatus can also be symmetrically
reversed, as long as the metric signature difference between $|X\rangle$
and $|T\rangle$ is opposite. This is because it is the relative signature
difference between them that holds physical meaning (keeping the speed
of light a constant). 

Since locally each basis vector can be seen as a fiber of the complementary
base subspace, the basis vectors of the two subspaces are locally
orthogonal
\begin{equation}
\langle T_{a}|X_{i}\rangle=0
\end{equation}

We can utilize the local metric $h_{\bar{i}j}$ of the subspace $\mathscr{H}_{X}$
to define the locally standard Levi-Civita connection and the intrinsic
Ricci curvature on $\mathscr{H}_{X}$, the non-vanishing components
are given by
\begin{equation}
\Gamma_{ij}^{k}=h^{k\bar{l}}\frac{\partial h_{j\bar{l}}}{\partial x^{i}},\quad\Gamma_{\bar{i}\bar{j}}^{\bar{k}}=h^{\bar{k}l}\frac{\partial h_{\bar{j}l}}{\partial x^{i*}},\quad R_{\bar{i}j}=-\frac{\partial\Gamma_{\bar{i}\bar{k}}^{\bar{k}}}{\partial x^{j}}=-\frac{\partial}{\partial x^{j}}\left(h^{\bar{k}l}\frac{\partial h_{\bar{k}l}}{\partial x^{i*}}\right)
\end{equation}
in which the indices of $h_{\bar{i}j}$ are uppered or lowered by
the metric of the same subspace, and we implicitly follow Einstein's
summation convention for tensors, summing over repeated upper and
lower indices and omitting the summation symbols.

The evolution of the intrinsic metric $h_{\bar{i}j}$ of the subspace
with respect to the ``external'' coordinates (time) outside the
subspace, from a geometric perspective, arises from the concept of
the extrinsic curvature of the subspace $\mathscr{H}_{X}$, describing
how the hypersurface $\mathscr{H}_{X}$ is bending with respect to
the external vector of the $\mathscr{H}_{T}$. By definition, the
extrinsic curvature is the derivative of the intrinsic metric $h_{\bar{i}j}$
with respect to its normal ``external'' coordinate $t^{a}$
\begin{equation}
K_{\bar{i}j}=\frac{1}{2}\frac{\partial h_{\bar{i}j}}{\partial t^{a}}|T^{a}\rangle
\end{equation}
which clearly describes the variation of the metric $h_{\bar{i}j}$
of the subspace $\mathscr{H}_{X}$ with respect to the state vector
(\ref{eq:T state}) of another entangled subspace $\mathscr{H}_{T}$.
We can also define two types of scalar extrinsic curvature by contracting
$K_{\bar{i}j}$ by $h^{\bar{i}j}$, 
\begin{equation}
K=h^{\bar{i}j}K_{\bar{i}j}=\frac{1}{2}h^{\bar{i}j}\frac{\partial h_{\bar{i}j}}{\partial t^{a}}|T^{a}\rangle,\quad K_{a}=\text{\ensuremath{\frac{1}{2}h^{\bar{i}j}\frac{\partial h_{\bar{i}j}}{\partial t^{a}}}}=\ensuremath{\frac{1}{2}}\ensuremath{\frac{\partial}{\partial t^{a}}\left(\ln\det h_{\bar{i}j}\right)}
\end{equation}
The derivative of the extrinsic curvature $K_{\bar{i}j}$ can be expressed
as a second-order derivative of the metric $h_{\bar{i}j}$, from $\langle T_{b}|T^{a}\rangle=-\delta_{b}^{a}$,
we have 
\begin{equation}
\frac{\partial K_{\bar{i}j}}{\partial t_{b}^{*}}\langle T_{b}|=\frac{1}{2}\frac{\partial^{2}h_{\bar{i}j}}{\partial t^{a}\partial t_{b}^{*}}\langle T_{b}|T^{a}\rangle=-\frac{1}{2}\frac{\partial^{2}h_{\bar{i}j}}{\partial t^{2}}
\end{equation}
Because the entangled state $|X,T\rangle$ is a standard quantum mechanical
state, the entire Hilbert space $\mathscr{H}_{X}\otimes\mathscr{H}_{T}$
in which it resides is Ricci-flat, satisfying the Ricci-flat K\"{a}hler-Einstein
equation
\begin{equation}
R_{\bar{I}J}(q_{\bar{I}J})=0\label{eq:Kahler-Einstein}
\end{equation}
where $q_{\bar{I}J}=h_{\bar{I}J}+s_{\bar{I}J}$ is the subspace decomposition
of the metric of the entire Hilbert space. When the Ricci-flat K\"{a}hler-Einstein
equation is decomposed into the $\mathscr{H}_{X}$ and $\mathscr{H}_{T}$
subspace, besides encompassing $N$ constraint equations, our current
focus lies on the $\frac{1}{2}N(N-1)$ relative evolution equations
between the basis vectors of the subspaces. More precisely, we are
interested in the equations that relate the intrinsic curvature $R_{\bar{i}j}$
of the $\mathscr{H}_{X}$ subspace to the extrinsic curvature $K_{\bar{i}j}$
\begin{equation}
-\frac{1}{2}\frac{\partial^{2}h_{\bar{i}j}}{\partial t^{2}}=R_{\bar{i}j}(h)+K(h)K_{\bar{i}j}(h)-2h^{k\bar{l}}K_{\bar{i}k}(h)K_{\bar{l}j}(h)\label{eq:evolution-h}
\end{equation}
which describes how the metric of $\mathscr{H}_{X}$ subspace evolves
with respect to the external vector of the $\mathscr{H}_{T}$ subspace.
And also analogous to the ADM decomposition of the Einstein equation,
the equation can also be seen as complex version of the ADM decomposition
of the Ricci-flat K\"{a}hler-Einstein equation. The geometric quantities
(intrinsic curvature and extrinsic curvature) on both sides of the
equation are covariant and no longer refer to any external absolute
parameters. Therefore, this equation is generally covariant and provides
an intrinsic description of the evolution of the quantum state basis
vectors $|X_{i}\rangle$ (or the metric $h_{\bar{i}j}$) of the entangled
subspace $\mathscr{H}_{X}$ relative to the state vector $|T\rangle$
(quantum clock) of the other entangled subspace $\mathscr{H}_{T}$. 

Symmetrically, we can also have an equation for $s_{\bar{a}b}$ 
\begin{equation}
\frac{1}{2}\frac{\partial^{2}s_{\bar{a}b}}{\partial x^{2}}=-\hat{R}_{\bar{a}b}(s)-\hat{K}(s)\hat{K}_{\bar{a}b}(s)+2s^{c\bar{d}}\hat{K}_{\bar{a}c}(s)\hat{K}_{\bar{d}b}(s)\label{eq:evolution-s}
\end{equation}
describing the metric of the clock subspace $\mathscr{H}_{T}$ evolves
with respect to the one-dimensional object's coordinate forming $\mathscr{H}_{X}$,
in which $\hat{R}_{\bar{a}b}(s),\hat{K}_{\bar{a}b}(s)$ and $\hat{K}(s)$
are the intrinsic Ricci curvature, extrinsic curvature and scalar
extrinsic curvature of $\mathscr{H}_{T}$, respectively. 

Let us examine the physical interpretation of the metric solutions
of the equations. During the instrument calibration prior to measurement,
mutually entangled quantum subsystems under-study $|X_{i}\rangle\in\mathscr{H}_{X}$
and the quantum instrument $|T_{a}\rangle\in\mathscr{H}_{T}$ are
prepared. The entangled state they form can be written as (\ref{eq:entangle state-2}).
During the measurement process following instrument calibration, there
is no further interaction between the quantum subsystem under-study
and the quantum instrument. The coefficients $C_{i}$ and initial
conditions for $h_{\bar{i}j}$ are established during the instrument
calibration process and do not change with respect to any (hypothetical)
external parameters afterward. The metric $h_{\bar{i}j}$ merely reflects
the relative changes between the quantum subsystem under-study $X$
and the quantum instrument $T$. When the quantum instrument is in
a certain quantum state $|T\rangle$ , the solutions to the equation
provide us with the metric $h_{\bar{i}j}$ or basis vectors $|X_{i}\rangle$
of the subsystem under-study that is entangled with its basis vectors
$|T_{a}\rangle$. Conversely, when the system is in a certain quantum
eigenstate $|X\rangle$, the equation (\ref{eq:evolution-s}) yields
the metric $s_{\bar{a}b}$ or basis vectors $|T_{a}\rangle$ that
is entangled with its basis vectors $|X_{i}\rangle$. Thus, by utilizing
the metric $s_{\bar{a}b}$ of eq.(\ref{eq:evolution-s}) or basis
vectors $|T_{a}\rangle$, we can perform corresponding conditional
projections on the entangled state $|X,T\rangle$ vector
\begin{equation}
\langle T|X,T\rangle=\sum_{i}C_{i}\langle T|T_{i}\rangle|X_{i}\rangle=\sum_{i}C_{i}t^{i*}|X_{i}\rangle=\sum_{i}\frac{C_{i}}{t_{i}^{*}}|X_{i}\rangle,\quad\left(\textrm{repeated}\:i\:\textrm{not}\:\textrm{summate}\right)
\end{equation}
in which the expressions $|X,T\rangle=\sum_{i}C_{i}|X_{i}\rangle|T_{i}\rangle$
and the conditional projection state $|T\rangle=\sum_{a}t^{a}|T_{a}\rangle$
are used. The coefficient $\frac{C_{i}}{t_{i}^{*}}$ preceding $|X_{i}\rangle$
represents the relative amplitude for the quantum subsystem under-study
in the state $|X_{i}\rangle$ when the quantum instrument is in a
certain conditional quantum state $|T\rangle$. Since $C_{i}$ is
determined by the initial conditions of instrument calibration and
remains fixed after the preparation of the entangled state, the evolution
equation (\ref{eq:evolution-h}) describes the evolution of the basis
vectors $|X_{i}\rangle$ with respect to $|T_{a}\rangle$. That is,
the solutions to the equation provide the correspondence between $|X_{i}\rangle$
and $t_{i}^{*}$, thereby establishing the relationship between the
relative amplitude $\frac{C_{i}}{t_{i}^{*}}$ and $t_{i}^{*}$.

Thus, in this theory, there is no concept of state vector evolution
as found in standard quantum mechanics (i.e., the unitary rotation
of state vectors with respect to external parameters). Here, the entangled
state vector $|X,T\rangle$ does not evolve (w.r.t. external parameter),
instead, it is the basis vectors $|X_{i}\rangle$ within the Hilbert
subspace that evolve relative to another subspace $|T_{a}\rangle$,
causing the projection $\langle T|X,T\rangle$ to appear as if $|X_{i}\rangle$
is evolving. Precisely because only the basis vectors undergo relative
evolution, this evolution is, in fact, universal or purely geometric.
We note that the evolution equation (\ref{eq:evolution-h}) or (\ref{eq:evolution-s})
does not depend on the system's mass nor explicitly contain the Planck
constant (just as the Berry phase does not explicitly involve the
Planck constant). The mass dependence and the Planck constant emerge
when the equation (\ref{eq:evolution-h}) or (\ref{eq:evolution-s})
recovers the dynamic Schr\"{o}dinger equation through the separation
of fast-changing and slow-changing parts in the basis vectors $|X_{i}\rangle$
(see next section). In this sense, the relative changes between basis
vectors on different subspaces described by the evolution equation
are purely geometric Berry-type phase changes, rather than being divided
into a dynamical part and a geometric (Berry-type) part with respect
to external parameter as the standard quantum mechanics has shown.

By employing the metric solution $s_{\bar{a}b}$ of eq.(\ref{eq:evolution-s})
to perform a partial trace of the density matrix $\rho_{X,T}=|X,T\rangle\langle X,T|$
over the $\mathscr{H}_{T}$ component (since the entangled state $|X,T\rangle$
is not in Schmidt decomposition form, the partial trace generally
requires contraction using the subspace metric rather than simply
summing the diagonal elements of the subspace), that is, by contracting
the density matrix with the metric of the subspace $\mathscr{H}_{T}$,
we obtain
\begin{equation}
\textrm{Tr}_{T}\left(\rho_{X,T}\right)=\langle T|X,T\rangle\langle X,T|T\rangle=\sum_{i,j}s_{\bar{i}j}C_{i}C_{j}^{*}|X_{i}\rangle\langle X_{j}|=\sum_{i}\frac{C_{i}C_{i}^{*}}{t_{i}^{*}t_{i}}|X_{i}\rangle\langle X_{i}|=\sum_{i}\left|\frac{C_{i}}{t_{i}}\right|^{2}|X_{i}\rangle\langle X_{i}|
\end{equation}
The diagonal elements of the matrix obtained after performing the
partial trace directly represent the relative probabilities (or conditional
probabilities) for the under-study subsystem in the state $|X_{i}\rangle$
when the quantum instrument is in the quantum state $|T\rangle$,
yielding $\textrm{P}\left(X_{i}|T\right)=\left|\frac{C_{i}}{t_{i}}\right|^{2}$.
As stressed before, the relative probability $\left|\frac{C_{i}}{t_{i}}\right|^{2}$
does not strictly equal to the standard absolute probability $\textrm{P}\left(X_{i}\right)=|x_{i}|^{2}=h_{\bar{i}i}(t)$
being a solution of eq.(\ref{eq:evolution-h}). Only in the separable
limit of the entangled state (\ref{eq:entangle state-2}) (or equivalently,
the quantum clock can be seen as a global external inertial clock),
it recovers the standard absolute probability $\textrm{P}\left(X_{i}|T\right)\overset{separable}{\approx}\textrm{P}\left(X_{i}\right)=\left\Vert |X_{i}\rangle\right\Vert ^{2}=h_{\bar{i}i}(t)$,
and hence, only in this sense, solving the evolution equation (\ref{eq:evolution-h})
for $|X_{i}\rangle$ or $h_{\bar{i}j}$ is approximately equivalent
to solve a Schr\"{o}dinger-like equation for the probability of the
particle's position $\textrm{P}\left(X_{i}|T\right)\overset{separable}{\approx}\textrm{P}\left(X_{i}\right)$
(we will see the eq.(\ref{eq:evolution-h}) recovering a Schr\"{o}dinger-like
equation in the next section). But in strict sense, the evolution
of the relative probability $\textrm{P}\left(X_{i}|T\right)$ must
be given by the evolution of $|T_{a}\rangle$ (or $s_{\bar{a}b}$)
w.r.t. $x$ according to eq.(\ref{eq:evolution-s}).

With the aid of this relative probability or the density matrix, the
quantum expectation value of an Hermitian mechanical operator $\mathbf{O}(X)$,
defined in $\mathscr{H}_{X}$ subspace, measured when the measuring
instrument is in the quantum state $|T\rangle$, can now be expressed
as
\begin{equation}
\textrm{Tr}_{T}\left[\rho_{X,T}\mathbf{O}(X)\right]=\sum_{i}\left|\frac{C_{i}}{t_{i}}\right|^{2}\langle X_{i}|\mathbf{O}|X_{i}\rangle
\end{equation}

Similarly, the converse can also be obtained. When the under-study
subsystem is in the quantum state $|X\rangle$, the relative probability
of the quantum instrument being in the state $|T_{i}\rangle$ is $\textrm{P}\left(T_{i}|X\right)=\left|\frac{C_{i}}{x_{i}}\right|^{2}$.
Correspondingly, the quantum expectation value of the mechanical operator
$\mathbf{O}(T)$, defined in $\mathscr{H}_{T}$, under this relative
probability is given by $\textrm{Tr}_{X}\left[\rho_{X,T}\mathbf{O}(T)\right]=\sum_{i}\left|\frac{C_{i}}{x_{i}}\right|^{2}\langle T_{i}|\mathbf{O}|T_{i}\rangle$.

We observe that the relative probabilities naturally emerge during
the process of taking partial traces over subsystems for the entangled
state $|X,T\rangle$. This differs from the partial tracing of entangled
states in Schmidt decomposition as presented in standard quantum mechanics.
Firstly, for entangled states in standard Schmidt decomposition, partial
tracing by the Kronecker's delta, the two reduced density matrices
obtained by taking partial traces over two distinct subsystems share
the same non-zero eigenvalues. In contrast, in our case, taking partial
traces over different subsystems by each non-trivial metric yields
different relative probabilities. Secondly, the subsystem basis vectors
of the entangled state in the Schmidt decomposition are orthonormal,
whereas the subsystem basis vectors of our prepared entangled states
are not orthonormal but instead define a non-trivial subspace metric.
This also leads to the fact that the metric of our subsystems satisfies
the general evolution equation (\ref{eq:evolution-h}) or (\ref{eq:evolution-s})
for local basis vectors, rather than being associated with the constant-curvature
Fubini-Study metric (which satisfies the constant-curvature Einstein
equation for complex projective spaces), as is the case for Schmidt's
entangled states.

\section{Recovering the Schr\"{o}dinger Equation}

In the previous section, we have seen that the evolution of $|X_{i}\rangle$
or $h_{\bar{i}j}$ w.r.t. the complementary external $t$ (\ref{eq:evolution-h})
is approximately equivalent to the evolution of $\textrm{P}\left(X_{i}|T\right)\overset{separable}{\approx}\textrm{P}\left(X_{i}\right)$
in the separable limit of the entangled state (\ref{eq:entangle state-2}),
therefore the evolution equations (\ref{eq:evolution-h}) or (\ref{eq:evolution-s})
plays a role analogous to that of the standard Schr\"{o}dinger equation
in describing the evolution of quantum states with respect to absolute
external parameters. Therefore, in the following, we will start from
the equation (\ref{eq:evolution-h}) and make some approximations
to return to the Schr\"{o}dinger equation limit. First, we linearize
the Ricci curvature as $R_{\bar{i}j}\approx-\frac{1}{2}\Delta_{x}h_{\bar{i}j}$,
where $\Delta_{x}$ is a linearized the Laplacian-Beltrami operator
with respect to $x_{i}$, then eq.(\ref{eq:evolution-h}) becomes
\begin{equation}
\frac{\partial^{2}h_{\bar{i}j}}{\partial t^{2}}=\Delta_{x}h_{\bar{i}j}+h^{k\bar{l}}\frac{\partial h_{\bar{i}k}}{\partial t^{a}}\frac{\partial h_{j\bar{l}}}{\partial t_{a}}-K_{a}\frac{\partial h_{\bar{i}j}}{\partial t_{a}}\label{eq:linearized equ-1}
\end{equation}
Using $h_{\bar{i}j}=\langle X_{i}|X_{j}\rangle$, and by equating
the corresponding terms on both sides of the equation, it can be transformed
into an equation in terms of the ket vector $|X_{i}\rangle$
\begin{equation}
\frac{\partial^{2}|X_{i}\rangle}{\partial t^{2}}+K_{a}\frac{\partial|X_{i}\rangle}{\partial t_{a}}=\Delta_{x}|X_{i}\rangle+h^{k\bar{l}}\frac{\partial h_{\bar{i}k}}{\partial t^{a}}\frac{\partial|X_{l}\rangle}{\partial t_{a}}\label{eq:linearized equ-2}
\end{equation}
If the ket vector $|X_{i}\rangle$ now has a fast-changing characteristic
frequency $\omega^{a}$ with respect to $t_{a}$, we will split $|X_{i}\rangle$
into two parts: a fast-changing part and a slow-changing part, i.e.
$|X_{i}\rangle=e^{-i\omega^{a}t_{a}}|e_{i}\rangle$, where $|e_{i}\rangle$
represents the slow-changing part compared to $\omega^{a}$, such
that $i\frac{\partial|e_{i}\rangle}{\partial t^{a}}\ll\omega_{a}|e_{i}\rangle$.
Then we obtain
\begin{equation}
\frac{\partial|X_{i}\rangle}{\partial t_{a}}=e^{-i\omega_{a}t^{a}}\left(\frac{\partial}{\partial t_{a}}-i\omega^{a}\right)|e_{i}\rangle\label{eq:d ket/dt}
\end{equation}
If we consider the scalar extrinsic curvature $K_{a}=\frac{1}{2}h^{\bar{i}j}\frac{\partial h_{\bar{i}j}}{\partial t^{a}}$
changes also very slowly
\begin{equation}
\frac{\partial}{\partial t^{b}}K_{a}\approx0,\quad\frac{\partial}{\partial t^{b*}}K_{a}\approx0\label{eq:K slow change}
\end{equation}
Since $K_{a}=\frac{\partial}{\partial t^{a}}\ln\sqrt{\det h_{\bar{i}j}}$,
the slow-changing approximation of this scalar extrinsic curvature
is equivalent to the local volume of the subspace, $\sqrt{\det h_{\bar{i}j}}$,
being approximately constant. This is a necessary condition for reverting
to the unitarity of the standard quantum mechanics. It is also equivalent
to the condition for returning to an inertial frame where no inertial
forces are present (see next section).

Now the left hand side of the equation becomes
\begin{equation}
\frac{\partial^{2}|X_{i}\rangle}{\partial t^{2}}+K_{a}\frac{\partial|X_{i}\rangle}{\partial t_{a}}\approx e^{-i\omega_{a}t}\left(\frac{\partial}{\partial t_{a}}-i\omega^{a}\right)\left(\frac{\partial}{\partial t^{a}}-i\omega_{a}+K_{a}\right)|e_{i}\rangle
\end{equation}
in which $\omega_{a}\omega^{a}=\omega^{2}$. When $|e_{i}\rangle$
is slow-changing, i.e. $\frac{\partial^{2}|e_{i}\rangle}{\partial t^{2}}\ll-2i\omega_{a}\frac{\partial|e_{i}\rangle}{\partial t_{a}}$,
the term $\frac{\partial^{2}}{\partial t^{2}}|e_{i}\rangle$ can be
neglected, the left hand side of the equation is approximately given
by
\begin{equation}
\frac{\partial^{2}|X_{i}\rangle}{\partial t^{2}}+K_{a}\frac{\partial|X_{i}\rangle}{\partial t_{a}}\approx e^{-i\omega_{a}t^{a}}\left[-2i\omega^{a}\left(\frac{\partial}{\partial t^{a}}+\frac{1}{2}K_{a}\right)-\omega^{2}\right]|e_{i}\rangle\equiv e^{-i\omega_{a}t^{a}}\left[-2i\omega^{a}\frac{D}{Dt^{a}}-\omega^{2}\right]|e_{i}\rangle\label{eq:left hand}
\end{equation}

Similarly, an independent equation satisfied by the bra vector $\langle X_{i}|$
can also be derived from the metric equation (\ref{eq:linearized equ-2}).
Therefore, the bra vector $\langle X_{i}|$ and the ket vector $|X_{i}\rangle$
are mutually independent state vectors. Assuming that $\langle X_{i}|$
varies slowly compared to $|X_{i}\rangle$,$\frac{\partial}{\partial t_{a}}\langle X_{i}|\ll-i\omega_{a}\langle X_{i}|$,
we have
\begin{equation}
\frac{\partial h_{\bar{i}k}}{\partial t_{a}}=\frac{\partial}{\partial t_{a}}\langle X_{i}|X_{k}\rangle=\frac{\partial\langle X_{i}|}{\partial t_{a}}|X_{k}\rangle+\langle X_{i}|\frac{\partial|X_{k}\rangle}{\partial t_{a}}\approx\langle X_{i}|\frac{\partial|X_{k}\rangle}{\partial t_{a}}=e^{-i\omega_{a}t^{a}}\langle X_{i}|\left(\frac{\partial}{\partial t^{a}}-i\omega_{a}\right)|e_{k}\rangle
\end{equation}
Since the slow-changing part of the ket vector satisfies $\frac{\partial}{\partial t_{a}}|e_{k}\rangle\ll-i\omega^{a}|e_{k}\rangle$,
we neglect the $\frac{\partial}{\partial t_{a}}|e_{k}\rangle$ term,
yielding
\begin{equation}
\frac{\partial h_{\bar{i}k}}{\partial t_{a}}\approx\left(-i\omega^{a}\right)h_{\bar{i}k}
\end{equation}
That is, the variation in the metric $h_{\bar{i}k}$ is approximately
dominated by the fast-changing part of the ket vector $|X_{k}\rangle$,
while $\langle X_{i}|$ serves as a slow-changing basis, which is
distinct from $\langle X^{i}|$, which is the fast-changing basis
dual to $|X_{i}\rangle$. This can also be seen from the fact that,
by $\langle X^{i}|X_{j}\rangle=\delta_{j}^{i}$,
\begin{equation}
\frac{\partial\langle X^{i}|}{\partial t_{a}}|X_{j}\rangle=-\langle X^{i}|\frac{\partial|X_{j}\rangle}{\partial t_{a}}\approx i\omega_{a}\delta_{j}^{i}\label{eq:d bra/dt}
\end{equation}
so $\langle X^{i}|$ is fast-changing. From alternative point of view,
since $\langle X^{i}|=h^{\bar{i}j}\langle X_{j}|$ and $\langle X_{j}|$
is now a slow-changing basis, the fast-changing property of $\langle X^{i}|$
is entirely attributed to the fact that the metric $h^{\bar{i}j}$
is fast-changing, specifically $\frac{\partial h^{\bar{i}j}}{\partial t_{a}}\approx\left(i\omega^{a}\right)h^{\bar{i}j}$.

Based on the above results, the right-hand side of equation (\ref{eq:linearized equ-2})
now approximately yields
\begin{equation}
\Delta_{x}|X_{i}\rangle+h^{k\bar{l}}\frac{\partial h_{\bar{i}k}}{\partial t^{a}}\frac{\partial|X_{l}\rangle}{\partial t_{a}}\approx e^{-i\omega_{a}t^{a}}\Delta_{x}|e_{i}\rangle+h^{k\bar{l}}\left(-i\omega^{a}\right)h_{\bar{i}k}e^{-i\omega_{a}t^{a}}\left(-i\omega_{a}\right)|e_{l}\rangle=e^{-i\omega_{a}t^{a}}\left(\Delta_{x}-\omega^{2}\right)|e_{i}\rangle\label{eq:right hand}
\end{equation}
Finally, equating the left hand side (\ref{eq:left hand}) and right
hand side (\ref{eq:right hand}), the eq.(\ref{eq:linearized equ-2})
can be approximately given by a Schr\"{o}dinger-like equation
\begin{equation}
i\frac{D|e_{i}\rangle}{Dt^{a}}\approx-\frac{1}{2\omega^{a}}\Delta_{x}|e_{i}\rangle\label{eq:schrodinger-like}
\end{equation}
in which 
\begin{equation}
\frac{D}{Dt^{a}}\equiv\frac{\partial}{\partial t^{a}}+\frac{1}{2}K_{a}\label{eq:covariant derivative}
\end{equation}
is a covariant derivative, in which the scalar extrinsic curvature
$K_{a}$ now plays the role of a connection for the parallel transport
of the slow-changing basis $|e_{i}\rangle$ on the Hilbert space $\mathscr{H}_{T}$.
If we regard the complex coordinates $t_{a}$ and $x_{i}$ in this
equation as real coordinates (mathematically, it is always possible
to appropriately choose coordinates or basis vectors $|T_{a}\rangle$
and $|X_{i}\rangle$ in the local quantum state subspace to render
the local coordinates $t_{a}$ and $x_{i}$ real), then, in terms
of the real coordinates, this equation formally becomes a Schr\"{o}dinger
equation.

To sum up, we have demonstrated that the equation (\ref{eq:evolution-h})
or its linearized approximations (\ref{eq:linearized equ-1}) or (\ref{eq:linearized equ-2})
describe the evolution of the metric $h_{\bar{i}j}$ or the basis
vectors $|X_{i}\rangle$ in the subspace $\mathscr{H}_{X}$ with respect
to the entangled state vector $|T\rangle$ in $\mathscr{H}_{T}$.
This is because we observe that the slow-changing part $|e_{i}\rangle$
of the basis vectors $|X_{i}\rangle$ satisfies an approximate ``free''
Schr\"{o}dinger evolution equation, where the fast-changing characteristic
frequency $\omega^{a}$ plays a role analogous to the mass $M=\hbar\omega^{a}$,
in which the mass and the Planck's constant emerge, and the potential
term in the Schr\"{o}dinger equation is embodied in the connection
(scalar extrinsic curvature $K_{a}$) of the covariant derivative.

\section{Non-Inertial Effects}

Because in this quantum theory, the evolution of quantum states is
no longer relative to the external Newtonian time of an absolute inertial
frame, but rather to the physical clock time in a general (non-inertial)
quantum reference system, and we have also observed that this ``free''
Schr\"{o}dinger evolution equation actually contains a potential term,
which generates an additional ``external force'', so this quantum
theory will automatically incorporate some effects of non-inertial
frames. As a result, certain ``inertial forces'' will automatically
emerge from the covariant derivative and the connection $K_{a}$.
This is analogous to the classical case that the Newton's law of motion
in inertial frame is generalized to the geodesic equation in general
non-inertial frames, and the inertial force (and gravity force) felt
by an object is encoded in the connection of the geodesic equation.

Through this connection $K_{a}$ in (\ref{eq:covariant derivative}),
if the state vector $|X\rangle$ moves along a path $c$ on $t\in\mathscr{H}_{T}$,
it acquires an additional non-integrable phase factor $e^{\varTheta}$,
(in general $\varTheta\in\mathbb{C}$)
\begin{equation}
|X^{\prime}\rangle=e^{\varTheta}|X\rangle\label{eq:exp(theta)}
\end{equation}
where 
\begin{equation}
\varTheta=\frac{1}{2}\int_{c}dt^{a}K_{a}\label{eq:complex phase}
\end{equation}
and $\varTheta$ not only encompasses the dynamics phase change (coming
from the dynamic Schr\"{o}dinger-like equation) of the slow-changing
basis $|e_{i}\rangle$ but also includes other additional phase changes
of the state $|X\rangle$. If the inner product of states is regarded
as a certain kind of distance, then factor $e^{\varTheta}\in\mathbb{C}$
can also be interpreted as the finite (complex) geodesic distance
between the quantum states $|X\rangle$ and $|X^{\prime}\rangle$.
When $c$ is a closed loop, i.e. when the state returns to its starting
point, the phase change contributed by the fast-changing dynamical
part in the phase disappears, leaving behind a phase change related
to the intrinsic geometry of the closed loop $c$. Since this closed
loop lies within the Hilbert subspace c rather than in a parameter
space, in this sense, the geometric phase is more akin to the Aharonov-Anandan
phase (a generalization of the Berry phase) induced by a closed loop
in the projected quantum state space. The phase $\varTheta$ and the
path accumulation of the potential term (the connection or extrinsic
curvature $K_{a}$) thus generally constitute a non-integrable phase.
The derivative of the ``potential'' $K_{a}$ gives rise to a certain
kind of ``external force'' 
\begin{equation}
F_{\bar{a}b}\equiv\frac{\partial K_{b}}{\partial t^{a*}}\label{eq:inertial force}
\end{equation}
where $F_{\bar{a}b}$ reflects a kind of \textquotedblleft inertial
force\textquotedblright{} that arises due to the choice of quantum
reference system when $|X\rangle$ is considered relative to $|T\rangle$.
Under the approximation (\ref{eq:K slow change}), we return to the
situation of an approximately inertial frame where it is inertial-force-free.

Furthermore, we can define another metric
\begin{equation}
g_{\bar{a}b}\equiv\left\langle \frac{\partial X^{i}}{\partial t^{a}}\right|\left.\frac{\partial X_{i}}{\partial t^{b}}\right\rangle \equiv G_{\bar{a}b}+\frac{i}{2}\Omega_{\bar{a}b}\label{eq:gab}
\end{equation}
in which $G_{\bar{a}b}$ and $\Omega_{\bar{a}b}$ are the real and
imaginary part of the metric. As a metric coming from the comparison
between $|X^{i}\rangle$ and $|T^{a}\rangle$, it can be interconverted
with the previous metrics $h_{\bar{i}j}$ through a tensor coordinate
transformation, for example,
\begin{equation}
g_{\bar{a}b}=\left\langle \frac{\partial X^{i}}{\partial t^{a}}\right|\left.\frac{\partial X^{j}}{\partial t^{b}}\right\rangle h_{i\bar{j}}=\left\langle \frac{\partial T_{a}}{\partial x_{i}}\right|\left.\frac{\partial T_{b}}{\partial x_{j}}\right\rangle h_{i\bar{j}}
\end{equation}
In contrast to the previous interpretation of $s_{\bar{a}b}$ and
$h_{\bar{i}j}$ as intrinsic metrics of the subspaces, $g_{\bar{a}b}$
can also be interpreted as a relative metric between $\mathscr{H}_{T}$
and $\mathscr{H}_{X}$, in certain sense, representing the pull-back
of the full-space metric $q_{\bar{I}J}$ onto the subspaces.

It can be shown that this metric $g_{\bar{a}b}$ is, in fact, equivalent
to the ``inertial force'' 
\begin{equation}
F_{\bar{a}b}=g_{\bar{a}b}\label{eq:F=00003Dg}
\end{equation}
From (\ref{eq:exp(theta)}), it can be observed that the real part
of the metric, $G_{\bar{a}b}=G_{b\bar{a}}=\textrm{Re}\left(g_{\bar{a}b}\right)$,
provides the (real) distance between two infinitesimally separated
state vectors $|X(t+dt)\rangle$ and $|X(t)\rangle$. Since the state
vector $|X\rangle$, as a local fiber in the non-trivial fiber bundle,
can only be normalized locally at $t$, both the length and the phase
of the state vector $|X\rangle$ will change relative to $|T\rangle$
as $t$ varies. Since 
\begin{equation}
|X_{i}(t+dt)\rangle=|X_{i}(t)\rangle+\left|\frac{\partial X_{i}(t)}{\partial t^{a}}\right\rangle dt^{a}+...
\end{equation}
so the real part $G_{\bar{a}b}$ corresponds to 
\begin{align}
2\textrm{Re}\left(d\varTheta\right) & =\left|\langle X^{i}(t+dt)|X_{i}(t)\rangle\right|^{2}-1=\textrm{Re}\left\langle \frac{\partial X^{i}}{\partial t^{a}}\right|\left.\frac{\partial X_{i}}{\partial t^{b}}\right\rangle dt^{a*}dt^{b}=G_{\bar{a}b}dt^{a*}dt^{b}\label{eq:Re(dTheta)}
\end{align}
Comparing the above equation and (\ref{eq:complex phase}), we see
the relation 
\begin{equation}
\textrm{Re}\left(d\varTheta\right)=\frac{1}{2}G_{\bar{a}b}dt^{a*}dt^{b}=\textrm{Re}\left(\frac{1}{2}K_{b}dt^{b}\right)\label{eq:Re(dTheta)-2}
\end{equation}
so we can conclude that, at certain approximation, the real part of
the metric is nothing but the real part of the ``inertial force''
\begin{equation}
G_{\bar{a}b}=\textrm{Re}\left(\frac{\partial K_{b}}{\partial t^{a*}}\right)=\textrm{Re}\left(F_{\bar{a}b}\right)
\end{equation}

Furthermore, if the real and imaginary parts of the metric are interconnected
by the complex structure operator $J$ of the state space (which satisfies
the eigenvalue equations $J\frac{\partial}{\partial t^{a}}=i\frac{\partial}{\partial t^{a}}$
and $J\frac{\partial}{\partial t^{a*}}=-i\frac{\partial}{\partial t^{a*}}$),
i.e.
\begin{equation}
JG_{\bar{a}b}=-\Omega_{\bar{a}b},\quad J\Omega_{\bar{a}b}=G_{\bar{a}b}
\end{equation}
so the imaginary part is given by 
\begin{equation}
\Omega_{\bar{a}b}=2\textrm{Im}\left(\frac{\partial K_{b}}{\partial t^{a*}}\right)=2\textrm{Im}\left(F_{\bar{a}b}\right)
\end{equation}
The imaginary part of the metric is also the imaginary part of the
``inertial force''. Therefore we have the relation eq.(\ref{eq:F=00003Dg}).

We observe that when the phase $\varTheta$ is integrable, or equivalent
speaking, the state space is K\"{a}hler, there exists a locally defined
Berry curvature $B_{\bar{a}b}$, which is defined as the exterior
derivative of the connection $K_{a}$, i.e. $B=dK$
\begin{equation}
B_{\bar{a}b}=\frac{\partial K_{b}}{\partial t^{a*}}-\frac{\partial K_{a}}{\partial t^{b*}}=F_{\bar{a}b}-F_{a\bar{b}}=F_{\bar{a}b}-F_{\bar{a}b}^{*}\label{eq:berry curvature}
\end{equation}
Comparing this equation with the definition $\Omega_{\bar{a}b}\equiv g_{\bar{a}b}-g_{\bar{a}b}^{*}$
and by using (\ref{eq:F=00003Dg}), we see that $\Omega_{\bar{a}b}$
can be interpreted as the Berry curvature (or field strength) $B_{\bar{a}b}$
\begin{equation}
\Omega_{\bar{a}b}\doteq B_{\bar{a}b}
\end{equation}
The symbol ``$\doteq$'' means that the equality holds only when
the K\"{a}hler form of the state space is closed, in other words,
the state space is strictly a K\"{a}hler manifolds or the phase $\varTheta$
is integrable (see below). However, in general, the geometry is only
almost K\"{a}hler but not strictly so. 

It is worth stressing that only when the phase $\varTheta$ is integrable,
(e.g. in a source-free case of the electromagnetic fields where the
field strength is closed $dB=0$), $d\Omega=0$, where $\Omega=\frac{i}{2}B_{\bar{a}b}dt^{a*}\wedge dt^{b}$
as the K\"{a}hler form of the state space is closed, in this case,
the state space is strictly K\"{a}hler.

Next, let us discuss some physical implications of the real and imaginary
parts of the non-integrable phase $\varTheta$. We observe that eq.(\ref{eq:Re(dTheta)})
can also be interpreted as an infinitesimal stretching of the length
of the state vector $|X\rangle$, which makes the evolution of $|X\rangle$
with respect to $|T\rangle$ no longer unitary in general. Consequently,
previously we could only derive the Schr\"{o}dinger equation with
unitary evolution under the assumption of slow-changing of the scalar
extrinsic curvature (\ref{eq:K slow change}) $G_{\bar{a}b}=\textrm{Re}\left(\frac{\partial K_{b}}{\partial t^{a*}}\right)\approx0$.
This is also an inevitable consequence of the relative evolution of
subsystems in entangled states, resulting from the quantum superposition
(quantum second-moment spreading \citep{Luo2015Dark,Luo:2023eqf,2024arXiv240809630L})
of the $T$ subsystem that serves as the reference frame.

This breakdown of unitarity can also be interpreted as an effect of
additional second-order fluctuations or extra broadening. Considering
the second-moment spreading of the effective energy $E$ corresponding
to the second-moment fluctuations of $t$, namely $\langle\delta E^{2}\rangle=\langle X^{i}|E^{2}|X_{i}\rangle-\langle X^{i}|E|X_{i}\rangle^{2}$, it leads to 
\begin{equation}
\left|\langle X^{i}(t+dt)|X_{i}(t)\rangle\right|^{2}=1-\langle\delta E^{2}\rangle dt^{2}\label{eq:1-<dE^2>dt^2}
\end{equation}
Therefore, comparing (\ref{eq:Re(dTheta)}) and (\ref{eq:1-<dE^2>dt^2}),
the real part of the metric $G_{\bar{a}b}$ is related to the second-moment
quantum fluctuations of $t$ or the second-moment fluctuations of
the effective energy, i.e. 
\begin{equation}
G_{\bar{a}b}=\langle\delta E_{a}\delta E_{b}\rangle\quad\textrm{or}\quad\left\Vert G_{\bar{a}b}\right\Vert \sim\langle\delta E^{2}\rangle
\end{equation}
where the negative sign of the difference is due to the timelike signature.
Hence, this ``inertial force'' $\left\Vert \textrm{Re}(F_{\bar{a}b})\right\Vert \sim\langle\delta E^{2}\rangle$
can also be regarded as arising from the quantum fluctuations of energy
or quantum fluctuations of the clock reference frame. We also observe
that the breakdown of unitarity caused by the second-order fluctuations
of the quantum clock and the broadening of spectral lines \citep{Luo:2023eqf}
both are coming from the additional ``inertial force'' effect. Conversely,
the effect of the ``inertial force'' always seems to be associated
with changes in the norm of the state vector or the breakdown of quantum
unitarity. This non-unitarity resulting from the broadening of the
quantum clock or spectral lines will be generalized to more general
quantum spacetime reference frames in \citep{Luo:2015pca,Luo:2019iby,Luo:2021zpi,2024arXiv240809630L},
and is the fundamental reason of coordinate transformation anomalies,
the cosmological constant, and gravity in general quantum spacetime
reference frames.

Similarly, since the imaginary part of the phase $\varTheta$ and
metric $g_{\bar{a}b}$, $\Omega_{\bar{a}b}$ is related to the Berry
curvature,
\begin{equation}
d\theta=\textrm{Im}\left(d\varTheta\right)=\frac{1}{4}\Omega_{\bar{a}b}dt^{a*}\wedge dt^{b}\doteq\frac{1}{4}B_{\bar{a}b}dt^{a*}\wedge dt^{b}\label{eq:Im(dTheta)}
\end{equation}
so the imaginary part describes an infinitesimal phase change of $|X\rangle$,
different from that the real part gives rise to the length change
of $|X\rangle$ (\ref{eq:Re(dTheta)}) and (\ref{eq:Re(dTheta)-2}).
The integration along the closed loop $c$ is given by
\begin{equation}
\theta[c]=\textrm{Im}\left(\varTheta\right)=\frac{1}{2}\textrm{Im}\oint_{c}dt^{a}K_{a}\doteq\frac{1}{4}\iint_{s}B_{\bar{a}b}dt^{a*}\wedge dt^{b}
\end{equation}
It describes the geometric phase change of the state vector $|X\rangle$
as it traverses a closed loop $c=\partial s$ on $\mathscr{H}_{T}$. 

To sum up, we see that the ``inertial force'' $F_{\bar{a}b}$ coming
from the potential $K_{a}$ in eq.(\ref{eq:schrodinger-like}) is
divided into two parts: the real part can be seen as providing a ``stretching
force'' $G_{\bar{a}b}$ related to the change in the length of the
state vector, while the imaginary part gives rise to a ``gauge force''
$B_{\bar{a}b}$ related to phase changes (we know that gauge fields
only alter the phase of the state vector without changing its magnitude).
The ``stretching force'' modifies the length of the state vector,
while the ``gauge force'' modifies its phase. In this sense, eq.(\ref{eq:schrodinger-like})
in general coordinate system automatically introduces a ``stretching
force'' and a ``gauge force'' at the quantum level. The profound
relationship between the imaginary part and the gauge force is also
quite intriguing, and can be found in other books discussing the geometric
phase in conventional quantum theory, as well as in the author's previous
articles. Our previous primary interest lies in the ``stretching
force'', because it actually manifests as an Einstein-gravity in
the quantum reference frame case \citep{Luo:2021zpi,2024arXiv240809630L}
being the generalization of the quantum clock. Generally speaking,
as long as a general complex connection (with both real and imaginary
parts) is introduced in the general transformation of the quantum
state relative to the quantum clock, resulting in a non-integrable
factor $e^{\varTheta}$ for the quantum state, it inevitably introduces
both gravity and gauge force through such a general coordinate transformations.
Whether this achieves some form of unification between gravity and
gauge force is a more profound question that lies beyond the scope
of this paper.

As observed earlier, the real part $G_{\bar{a}b}$ describes the second-order
fluctuations in the length of vectors on $\mathscr{H}_{X}$ relative
to those on $\mathscr{H}_{T}$, reflecting the degree of unitarity
violation associated with changes in the norm of the state vector.
If we note that the scalar extrinsic curvature actually describes
the first-order variation of the local volume element of $\mathscr{H}_{X}$
relative to the vector $|T\rangle$,
\begin{equation}
K_{b}=\frac{\partial}{\partial t^{b}}\ln\sqrt{\det h_{\bar{i}j}}
\end{equation}
so the real part of the metric can be given by
\begin{equation}
G_{\bar{a}b}=\textrm{Re}\left(\frac{\partial K_{b}}{\partial t^{a*}}\right)=\textrm{Re}\left(\frac{\partial^{2}}{\partial t^{a*}\partial t^{b}}\ln\sqrt{\det h_{\bar{i}j}}\right)\label{eq:Re(g) from Kahler potential}
\end{equation}
It naturally follows that the (logarithmic) magnitude of the local
volume element of $\mathscr{H}_{X}$ undergoes second-order variations
with respect to $|T\rangle$. The changes in the local volume of $\mathscr{H}_{X}$
are precisely caused by the quadratic correction to the length of
its quantum state due to second-order fluctuations. Under the slow-changing
approximation of the extrinsic curvature (\ref{eq:K slow change}),
the local volume is approximately conserved, and unitarity approximately
holds. Meanwhile, the imaginary part $\Omega_{\bar{a}b}$, acting
as the Berry curvature, describes the second-order phase change of
the state vector on $\mathscr{H}_{X}$ relative to that on $\mathscr{H}_{T}$.

We observe from (\ref{eq:Re(g) from Kahler potential}) that the logarithm
of the local volume element of $\mathscr{H}_{X}$, namely $\ln\sqrt{\det h_{\bar{i}j}}$,
formally plays a role analogous to the K\"{a}hler potential, meaning
that the complex metric can be expressed in terms of a single volume
form
\begin{equation}
g_{\bar{a}b}=\partial_{\bar{a}}\partial_{b}\ln\sqrt{\det h_{\bar{i}j}}\label{eq:g from Kahler potential}
\end{equation}
If the phase $\varTheta=\frac{1}{2}\int_{c}dt^{a}K_{a}$ is integrable
(independent to the path $c$), meaning that the local Berry curvature
$B=B_{\bar{a}b}dt^{a*}\wedge dt^{b}$ is closed ($dB=d^{2}K=0$),
and so if this local Berry curvature corresponds to the imaginary
part $\Omega_{\bar{a}b}$ of the local metric, the quantum state space
equipped with this Hermitian complex metric must be a K\"{a}hler space.
In this case, the gauge field strength 2-form $\frac{i}{2}B_{\bar{a}b}dt^{a*}\wedge dt^{b}\doteq\frac{i}{2}\Omega_{\bar{a}b}dt^{a*}\wedge dt^{b}$
serves as the K\"{a}hler form, reflecting the symplectic structure
of the K\"{a}hler space. A K\"{a}hler space is one that simultaneously
possesses a Riemannian structure, a symplectic structure, and is compatible
with its complex structure. When the quantum state space is a K\"{a}hler
space, or equivalently, the phase $\varTheta$ is integrable, due
to the enhanced symmetry of the K\"{a}hler metric, which can be represented
by a single K\"{a}hler potential, many problems by solving the multi-component
Ricci-flat K\"{a}hler-Einstein equations can be reduced to solving
a single complex Monge-Amp\`{e}re equation, which simplifies the problem
of solving the K\"{a}hler-Einstein equations. When there exists a
constant volume form $\sqrt{\det h_{\bar{i}j}}$ of state subspace
(w.r.t. another subspace), i.e. a constant K\"{a}hler potential $\ln\sqrt{\det h_{\bar{i}j}}$,
this degenerates to the case of a unitary quantum mechanics, so the
state space of the unitary standard quantum mechanics is K\"{a}hler,
which is suggested by literature e.g. \citep{Ashtekar:1997ud,Bengtsson:2006rfv}.
In particular, when the curvature of its subspace is proportional
to the metric, $R_{\bar{a}b}=(N+1)g_{\bar{a}b}$, the quantum state
space is homeomorphic to the maximally symmetric sphere $S^{2N+1}$
and this K\"{a}hler metric $g_{\bar{a}b}$ is known as the Fubini-Study
metric \citep{Pandya:2006mrv,Matassa:2020dnc}. However, the subspace
metric $h_{\bar{i}j}$ of interest to us generally does not possess
constant curvature or a constant local volume form, so the subspace
metric of the quantum state here is generally not Fubini-Study in
this framework. Nonetheless, in general, our phase $\varTheta$ is
not strictly integrable (depending on the path and history), and the
Berry curvature can no longer be locally defined. In this case, the
quantum state subspace of the framework is not strictly K\"{a}hler
but can be regarded as an almost K\"{a}hler manifold.

\section{Comparison with Some Past Works}

Given the long-standing research on the relational formulations and
internal evolution of quantum mechanics, coupled with the already
vast body of literature on the subject, it has become impossible for
us to fully grasp all the existing works. While our research may have
been partially or indirectly inspired by some past studies in terms
of major philosophical directions, the core ideas of our work are
independently discovered. Our primary motivations for developing the
theory of quantum reference frames are entirely different from those
in previous literature. The initial few papers were motivated by the
cosmological constant problem \citep{Luo2014The,Luo2015Dark,Luo:2015pca}
and quantum gravity, which must involve quantum spacetime as reference
frame, whereas the present paper focuses solely on quantum mechanics
which involves only quantum clock. Nevertheless, it was only later
that we realized our thoughts converged with those of these past studies
in certain aspects, albeit through different paths.

Our paper achieves results that are similar in many respects to those
in the work by Page and Wootters \citep{PhysRevD.27.2885}. However,
there are also numerous differences between our paper and their study.
Firstly, in Page and Wootters' article, the non-zero Hamiltonian of
the clock subsystem still exists, and evolution of the density matrix
of the quantum clock is realized through it. In contrast, our formulation
is purely covariant and no longer incorporates a Hamiltonian, since
the presence of any non-zero Hamiltonian implicitly implies a preferred
inertial frame which is a concept that the general relativity aims
to discard from the outset. Therefore, the quantum clock reference
system discussed in our article is a general one, i.e. any general
quantum system can be seen as a quantum clock, where inertial frames
hold no special status whatsoever, while Page-Wootters's theory still
presupposes a particular (inertial) type of internal clock. That is
the reason why the general non-inertial effects appear which seem
absence in Page-Wootters' theory. We also observe that it is precisely
the quantum fluctuations of the quantum reference system that dictate
the non-existence of rigorous inertial frames at the quantum level.
Secondly, although both our paper and theirs provide methods for calculating
conditional probabilities, these methods are entirely distinct. In
their work, the calculation of conditional probabilities is carried
out by taking the conventional partial trace over the clock (sum over
the diagonal elements) which is equivalent to a projection of the
quantum state onto a specific quantum clock reference frame subspace
which is flat. In other words, this quantum reference frame is still
associated with some kind of flat inertial frame, rather than being
a general quantum coordinate system. In our framework, conditional
probability are computed via partial traces over the clock by contracting
by a generally curved metric of the quantum clock reference frame
subspace. It is also noted that in Page-Wootters' calculation of the
conditional probability, taking the conventional partial trace and
divided by the probability of the quantum clock are two separate steps.
However, in our approach, these two steps are synthesized into a single
step: the contracting by the clock metric, which features a more natural
and covariant geometric construction. Moreover, the entangled state
(\ref{eq:entangle state-2}) prepared during calibration, is fixed,
and does not evolve with respect to external Hamiltonian, instead,
the relative evolution of the under-study $|X_{i}\rangle$ with respect
to the clock subsystem arises solely from the projection metric $s_{\bar{a}b}\neq\delta_{\bar{a}b}$
of the subspace which only evolves internally. And through geometric
method, the paper has found the dynamical equation governing the evolution
of the general metric (\ref{eq:evolution-s}), which indirectly gives
the time evolution of the conditional probability by the internal
evolution of the subspace metric, completely different from Page-Wootters
and related theory \citep{PhysRevD.27.2885,Foti:2020erm,Woods2019pagewootters,qfns-48vq}.
And moreover, in this framework, it is more transparent to seen the
relation between the conditional probability $\left|\frac{C_{i}}{t_{i}}\right|^{2}$
and the absolution probability $|x_{i}|^{2}$, that is, only in the
separable limit of the entangled state (\ref{eq:entangle state-2}),
or equivalently, the quantum clock can be seen as a global external
inertial clock, these two probabilities are equal, which is not easily
to seen in the Page-Wootters' original and subsequence papers. As
a consequence, it gives us a deeper understanding to Kuchar's criticism
about the ``wrong propagator'' in the Page-Wootters' theory \citep{Kuchar:1991qf}:
that is, the Dirac delta propagator $\delta(x-x^{\prime})\delta(t-t^{\prime})$.
The delta propagator is, of course, a correct propagator or conditional
amplitude for the general covariant system as a whole without any
external evolving parameter (particularly, it is this propagator shows
the mass-independence as the quantum equivalence principle \citep{2024arXiv240809630L}
has pointed out). To crossover to the well-known standard (mass-dependent)
propagator, the clock must be separable as an unentangled direct product
state, and hence the state of the clock can be factored out as an
global external inertial clock in which process a Hamiltonian or mass-dependence
emerges in the standard propagator and the conditional probability
equals to the standard absolute one. To the best of our knowledge,
this resolution to the Kuchar's criticism is also not explicitly claimed
before. For Kuchar's another criticism of the Page-Wootters mechanism
regarding the issue that the conditional probability may violate the
Hamiltonian constraint, and more specifically, the problem of whether
the clock projection operator be non-commutative with the Hamiltonian
constraint operator. The problem also does not exist within our framework.
Comparing with the \textquotedblleft projection interpretation\textquotedblright{}
of the conditional probability in Page-Wootters's paper, it is more
transparent to see that the ``projection'' by the clock metric $s_{\bar{a}b}$
is just a ``coordinate'' transformation (or gauge transformation)
of the entangled state (which satisfies the Hamiltonian constraint)
in the full Hilbert space, choosing a specific clock metric is just
equivalent to choosing a specific ``coordinate system'', the conditional
probability is a coordinate system fixed and reduced version of invariant
inner product. In other words, the clock metric as an induced subspace
metric from the full space naturally satisfying the Ricci flat K\"{a}hler-Einstein
equation of the full space and hence the conditional probability naturally
obeys the constraint. It is also for this general covariant reason
of the conditional amplitude of the framework, Kuchar's another criticism
of ``Inappropriate for the relativistic Klein-Gordon systems'' (this
criticism is also originated from the non-general-covariance of Page-Wootters'
internal clock time) does not exist in this framework as well, because
there must be a ``coordinate'' transformation to relate the general
covariant internal clock time in this framework to an external relativistic
Minkovski time for Klein-Gordon systems. To sum up, this article is
not merely a simple re-formulation of the Page-Wootters' framework,
rather, it can be regarded as presenting a more general generalization,
the new physics are the non-inertial effects coming from the general
covariance without a Hamiltonian, just as that general relativity
is not a simple re-formulation of Newton's laws, the new physics is
the gravity naturally incorporated.

Although our discussions only focus on the quantum clock (not a quantum
spacetime), it is of course, in essence, a quantum reference frame
theory. However, frankly speaking, the results obtained in the existing
literature on quantum reference frames, (e.g. \citep{PhysRevD.30.368,Rovelli:1990pi,RevModPhys.79.555,Giacomini:2017zju,Loveridge:2017pcv,2020Quant...4..225V,Hohn_2020,Giacomini:2020ahk}
and references therein) are still quite preliminary, in the sense
that it is still a long way off from the real goal of this subject:
quantum gravity. Compared our works with the existing literature of
quantum reference frame, firstly, most current articles on quantum
reference frames are still based on the description of quantum reference
frames via non-zero Hamiltonian. As previously mentioned, this implicitly
implies the selection of a preferred inertial frame, making it difficult
for these theories to derive non-inertial effects, let alone naturally
incorporate gravitational effects. And so, the coordinate transformations
discussed in these literature can only be seen as transformations
between quantum inertial frames, not a general quantum coordinate
transformation. On the other hand, current literature on quantum reference
frames still lack a fundamental quantum equivalence principle as a
guiding principle to re-interpret the universal quantum properties
of material quantum reference frames as the universal properties of
the abstract concept of quantum time itself or quantum spacetime itself.
In our paper on the quantum equivalence principle \citep{2024arXiv240809630L},
we first point out that only non-dynamical universal part of quantum
second-order moment fluctuation (not coming from its Hamiltonian or
mass, but coming from a cutoff scale dependent broadening of the Dirac
delta propagator mentioned in the previous paragraph) of physical
reference frames, can universally describe the second-order moments
of quantum clocks or quantum spacetime, thereby reconciling the inherent
contradiction between the mass-dependence of the Schr\"{o}dinger equation
of standard quantum mechanics and the equivalence principle (different
from most of the literature on the subject, e.g. \citep{Giacomini:2020ahk,Das:2023cfu}).
Therefore, our quantum reference frame theory is built upon a covariant
theory with a zero Hamiltonian, which is different from most of the
literature of the subject. Consequently, in this paper, such internal
quantum clocks naturally give rise to non-inertial effects. Furthermore,
getting back to our main motivation of the paper: a quantum gravity,
(quantum clock discussed in the article is used as a conceptual illustration
example), we would like also discuss here some comparisons between
our quantum spacetime reference frame theory (e.g. \citep{Luo:2021zpi,2024arXiv240809630L,2024chinaxiv})
beyond the scope of quantum clock and the existing literature on the
subject. In our quantum spacetime reference frame (a non-linear sigma
model with zero Hamiltonian as a generalization of the quantum clock,
which is more suitably treated by functional integral approach rather
than the canonical relative state approach), the spacetime diffeomorphism
anomalies naturally lead to the emergence of gravity (see \citep{Luo:2021zpi,2023AnPhy.45869452L,2024arXiv240809630L}),
and the cancellation of anomalies is connected to the cosmological
constant---a feat that has not been truly achieved in the most literature
on the quantum reference frame. Secondly, our quantum reference frame
theory provides the renormalization group flow of the quantum spacetime
(which is, to an approximation, a Ricci flow) and its associated monotonic
entropy functional \citep{Luo:2022statistics,Luo:2025cer}, which
gives an origin of spacetime and gravitational entropy, and enabling
a global control of a gravitational systems (since a covariant system
like gravity lacks a global time coordinate and hence the concept
of energy/Hamiltonian no longer provides a global control like usual
dynamical system), a feat not successfully achieved in most current
quantum reference frame literature. Moreover, a general quantum coordinate
transformation is given by a Jacobian in the functional integral approach
which leads to spacetime diffeomorphism anomaly and gravity, or equivalently,
given by an integral transformation via a heat kernel of the conjugate
heat flow equations which is coupled to the spacetime Ricci flow.
The resulting general quantum coordinate transformation is non-unitary
due to the non-inertial effects (including the diffeomorphism anomaly,
non-isometric of the Ricci flow and gravity, etc.) coming from the
second-order moment quantum fluctuations, which is completely different
from most of the existing literature on the subject.

Of course, as a quantum reference frame theory, the philosophical
interpretation of quantum mechanics presented in this paper is much
closer to that of the relational quantum mechanics \citep{Rovelli:1995fv,Rovelli:2021thu},
and has little to do with the ``relative state'' in the so-called
many-worlds interpretation, despite the semantic similarity between
``relational'' and ``relative''---we do not wish to get bogged
down in semantic quibbles here, and we also do not want to get entangled
in the philosophical disputes among various schools of thought derived
from the relational interpretation. The fundamental viewpoint we advocate
in this paper is that quantum states are relative (w.r.t. quantum
apparatus), and physical properties can only be defined relatively.
Hence, there is no global section (of the non-trivial fiber bundle)
serving as a global wave function and moreover no global underlying
``hidden variable''. The ``relative state'' is a local section
on a non-trivial fiber bundle, whereas the many-worlds interpretation
still posits a global wave function, with observers merely perceiving
branches of this global wave function. Of course, unlike relational
quantum mechanics, which merely offers a new interpretation of the
existing mathematical formalism of quantum mechanics, we strive to
more thoroughly incorporate the spirit of general relativity into
the framework of quantum mechanics.

\section{Conclusions}

We summarize the main conclusions of this paper as follows:

(1) This paper presents a quantum theory based on entangled or relative
states rather than absolute quantum states. This quantum theory describes
how a quantum system evolves relative to another quantum system, especially,
describing a quantum state of the relationship between the quantum
under-study system and that of the quantum measuring apparatus, without
relying on any external absolute parameter. The relative interpretation
serves as a physical interpretation of the entangled state, and the
geometrical interpretation of the entangled state is in fact a non-trivial
fiber bundle. The picture of the non-trivial fiber bundle interpretation
of entangled state also constitutes an important motivation for our
proposal that a locally curved state space geometric description is
required to depict the relative evolution of quantum states.

(2) We gives a non-trivial fiber bundle interpretation to the entangled
state $|X,T\rangle=\sum_{\tau}C_{\tau}|X(\tau)\rangle\otimes|T(\tau)\rangle$,
in which $|T(\tau)\rangle$ is the local base space, $|X(\tau)\rangle$
is the fiber over the local base $|T(\tau)\rangle$, the entangled
state $|X,T\rangle$ is then the fiber bundle, and the projection
$T^{-1}:|X,T\rangle\rightarrow|X(\tau)\rangle$ from the fiber bundle
to a section of the fiber gives rise to the relative amplitude of
the under-study quantum state $|X(\tau)\rangle$ relative to the clock
state $|T\rangle$. So there is a relative probability interpretation
concerning the probability of state $|X(\tau)\rangle$ relative to
the measuring apparatus state $|T\rangle$, rather than an absolute
probability interpretation relative to external parameters. The relative
or conditional probabilities derived from an entangled state reflect
that our information about the entangled system is incomplete. When
one subsystem of an entangled state learns, via a classical channel,
the measurement result of the other subsystem, it can utilize this
additional information to update our knowledge about this local subsystem
in the sense of the Bayesian theorem. Only when the entangled state
is separable or the measuring apparatus can be considered classical,
then entangled state degenerates into a direct product state (the
non-trivial fiber bundle degenerates to a trivial bundle), does this
relative probability recovers the absolute probability of the standard
quantum mechanics.

(3) The entangled state of the under-study state $|X\rangle$ and
the clock apparatus state $|T\rangle$, prepared during calibration,
does not evolve with respect to external parameters. However, the
relative evolution of the under-study quantum state $|X\rangle$ relative
to the clock state $|T\rangle$ in the intrinsic quantum theory can
be described by an evolution equation of the total quantum state space
coming from the Ricci-flat K\"{a}hler-Einstein equation. The evolution
equation is a generally covariant equation that no longer refers to
any external absolute parameters and constrained by a vanishing Hamiltonian,
but only relies on internal relative descriptions. In this sense,
this framework achieves a geometric formulation of a quantum theory
describing the under-study quantum system relative to the quantum
apparatus. This equation has a linear and non-relativistic approximation
to the Schr\"{o}dinger equation.

(4) From this framework, the evolution of quantum states is now the
evolution of basis vectors of one subspace relative to those of another
subspace. Although the entangled state (relative to external parameters)
does not evolve, the projection components of the entangled state
onto the basis vectors of the subspace evolve relative to the clock
subspace. In this picture, there is no concept of a state vector evolving
relative to unchanging basis vectors; instead, the entangled state
vector remains unchanged while the basis vectors change. Consequently,
there is no longer a distinction between quantum dynamical phases
(where basis vectors remain unchanged and state vectors change) and
geometric phases (where basis vectors change relative to external
parameters). Now, the ``dynamical phase'' essentially also arises
from the geometric phase due to changes in the basis vectors. In this
sense, the quantum dynamical evolution is geometrized, as the Newton's
dynamical law is geometrized by the geodesic equation in general coordinate
system.

(5) Within this framework, there is essentially no difference between
the time evolution of the under-study system (by a quantum clock)
and a measurement process of other aspects of the under-study system
(by other quantum instruments). The distinction lies in the fact that
the quantum clock provides a reference state with no interaction with
the state of the under-study system, whereas other general measuring
instruments offer other reference states with certain interaction
with the state of the under-study system. Consequently, in certain
sense, this framework offers a unified description that bridges the
time evolution (which is described by the unitary Schr\"{o}dinger
equation in standard quantum mechanics) and the general quantum measurement
process (which is depicted by a non-unitary collapse of wavefunction
in the standard quantum mechanics).

(6) For the intrinsic quantum system of the measured quantum state
$|X\rangle$ relative to the clock state $|T\rangle$, we observe
the natural emergence of ``inertial forces'' associated with the
choice of a general clock reference state. This ``inertial force''
$F_{\bar{a}b}$ is related to the non-trivial relative metric $g_{\bar{a}b}$
between the under-study quantum state $|X\rangle$ and the clock apparatus
state $|T\rangle$. The ``inertial force'' and its quantum effects,
derived from the general choice of reference frame, are issues of
concern in \citep{Luo:2021zpi}. The real part of this ``inertial
force'' brings about certain ``stretching force'' and second-order
broadening of the entangled subsystem which leads to the destruction
of quantum state unitarity and anomaly (volume changes in the subsystem
quantum state space and information loss dependent on history), while
the imaginary part brings about certain ``gauge force'' and geometric
phases. When the phase factor is non-integrable, just as the geometric
phase brought about by the imaginary part is history-dependent, the
destruction of unitarity by the real part is also history- or path-dependent.
This historical dependence of quantum non-unitary effects has already
been observed in quantum field theory in curved spacetime. Throughout
this paper, we will see that the quantum unitarity-violation effects
and anomalies brought about by non-inertial frames are very common
and general conclusions. However, the ``inertial force'' $F_{\bar{a}b}$
derived in this simple system does not yet exhibit gravitational characteristics.
It is merely a general inertial force. The reason is that our current
system is still a simple (single reference degree of freedom) quantum
clock reference system. When extended to a multiple reference degrees
of freedom, namely, a quantum spacetime reference system, the characteristics
of an Einstein-gravity arising from the unitarity-violation and anomalies
caused by this type of inertial force will become apparent (e.g \citep{Luo:2021zpi,2024arXiv240809630L}),
it automatically leading to a gravity theory coming from the general
covariance of quantum reference systems. Our previous works on the
functional integral approach to the quantum spacetime reference frame
only focuses on the real part and its relation to the anomaly and
gravity, while the effects of the imaginary part have not been taken
into account, now it seems that the relative state approach also reveals
the effects of the imaginary part and its relation to the Berry phase
and ``gauge force''.

(7) The relative evolution equation for the quantum state of the one-dimensional
object's position relative to a single-degree-of-freedom quantum clock
reference system has already been a high dimensional evolution equation.
Therefore, it can be imagined that if more degrees of freedom of the
quantum spacetime reference system are involved, this canonical framework
will become even more complicated. This canonical framework, based
on differential equations for quantum state evolution, is more suitable
for conceptually demonstrating its relationship with the standard
quantum mechanical Schr\"{o}dinger equation and its probability interpretation,
helping us observe more effects that are not easily seen through the
functional integral approach. On the other hand, the action and functional
integral quantization framework is actually more amenable to practical
mathematical calculations and to study quantum gravity. Each method
has its own advantages and disadvantages. To facilitate comparison
between these two methods, we will provide the functional method for
a single-degree-of-freedom quantum clock in the Appendix for easy
comparison. For the functional treatment of quantum spacetime reference
systems with more degrees of freedom, please refer to \citep{Luo:2021zpi,2024arXiv240809630L,2024chinaxiv}.
\begin{acknowledgments}
This work was supported in part by the National Science Foundation
of China (NSFC) under Grant No.11205149, and the Scientific Research
Foundation of Jiangsu University for Young Scholars under Grant No.15JDG153.
\end{acknowledgments}

\section*{Appendix: Functional Integral Approach}

Previously, we primarily examined the relationship between the quantum
state of the under-study system and that of the measuring apparatus
from the perspective of quantum states, particularly entangled states
or relative state. We discussed the evolution equation that governs
the evolution of the quantum state space metric of the entangled subsystem
relative to another subsystem, as well as its relationship with the
Schr\"{o}dinger equation. Additionally, we explored how to project
entangled states onto subspaces using subspace metrics. This approach
represents a canonical formulation based on quantum states. In this
appendix, we will address this issue from the perspective of relative
evolution based on the functional method. The advantage of the functional
integral approach is that quantum operators, treated as q-numbers
(quantum numbers), can be regarded as ordinary c-numbers (classical
numbers) when subjected to mean-field or semi-classical approximation.
This facilitates a recovering to the conventional standard quantum
theory, making it easier to compare with the standard quantum theories
and to discern how this theory surpasses the standard framework.

As previously mentioned, what holds physical meaning is the relationship
between the object's position and the position of the clock pointer.
In order to clarify how the position operator $X$ of the object evolves
with respect to the position operator $T$ of the clock pointer, we
consider the action formulation of this theory
\begin{equation}
S_{X}=\int d\tau\left[\frac{1}{2}m_{X}\left(\frac{dX}{d\tau}\right)^{2}-V(X)\right],\quad S_{T}=\frac{1}{2}m_{T}\int d\tau\left[\left(\frac{dT}{d\tau}\right)^{2}\right]\label{eq:SX and ST}
\end{equation}
Unlike the relative state approach, which does not require any external
absolute parameters, the functional method now necessitates a global
parameter $\tau$, but the physical results do not depend on it. Both
the object's position $X$ and the clock $T$ evolve relative to this
global parameter $\tau$, and in this way, $X$ evolves relative to
$T$. In this model, the global parameter $\tau$ can be interpreted
as the ``proper time'' of the laboratory as discussed in section
II.

The action for the object is given by the kinetic energy $\frac{1}{2}\left(\frac{dX}{d\tau}\right)^{2}$
minus certain potential energy $V(X)$ with respect to the laboratory
proper time $\tau$. We assume that the clock pointer moves uniformly
with respect to the laboratory time $\tau$, so the action for the
clock consists solely of kinetic energy. For the two mutually entangled
systems $|X,T\rangle$ has no interaction with each other, the actions
for the object $X$ and the clock $T$ are the direct sums of the
actions of two non-interacting systems,
\begin{equation}
S[X,T]=S_{X}+S_{T}=\int d\tau\left[\frac{1}{2}m_{X}\left(\frac{dX}{d\tau}\right)^{2}-V(X)+\frac{1}{2}m_{T}\left(\frac{dT}{d\tau}\right)^{2}\right]
\end{equation}
We observe that, since the total action $S[X,T]$ does not contain
a linear term in $T$, the total average energy of the system w.r.t.
the internal clock time $T$ is zero
\begin{equation}
\frac{\delta S}{\delta T}=\langle E\rangle=0\label{eq:<E>=00003D0}
\end{equation}
Thus, the total Hamiltonian $H_{X}+H_{T}$ of the entire system w.r.t.
the internal clock time equals zero, and the Schr\"{o}dinger equation
governing the evolution of the system's quantum state $|X,T\rangle$
with respect to $T$ is, strictly speaking, a Wheeler-DeWitt equation
\begin{equation}
\left(H_{X}+H_{T}\right)|X,T\rangle=0\label{eq:WD equ}
\end{equation}

This equation corresponds to the fact that the entangled state does
not evolve w.r.t. $T$ and remain fixed after calibration. Instead,
the evolution of the object's position $X$ relative to the quantum
clock $T$ arises because both $X$ and $T$ both evolve with respect
to the common laboratory proper time $\tau$. Therefore, to obtain
the evolution of $X$ relative to the clock $T$, we only need to
perform a semiclassical coordinate transformation $\tau\rightarrow T$
(``semiclassical'' means that the operator $T$ is treated using
the mean-field/semiclassical approximation, handled as a c-number
rather than a q-number and ignore its fluctuation), and then the action
can be rewritten as
\begin{equation}
S[X,T]=\int dT\left\Vert \frac{d\tau}{dT}\right\Vert \left\{ \left(\frac{dT}{d\tau}\right)^{2}\left[\frac{1}{2}m_{X}\left(\frac{\delta X}{\delta T}\right)^{2}+\frac{1}{2}m_{T}\right]-V(X)\right\} 
\end{equation}
in which $\left\Vert \frac{d\tau}{dT}\right\Vert $ represents the
lapse function of time, which is essentially the Jacobian of the coordinate
transformation $\tau\rightarrow T$. This transformation formally
converts the original integral with respect to $\tau$ into an integral
with respect to $T$. Quantizing this theory is equivalent to calculating
the partition function
\begin{equation}
Z=\int\mathscr{D}X\mathscr{D}Te^{iS[X,T]}
\end{equation}
By employing the semiclassical approximation, we only consider the
semi-classical (S.C.) value of the clock $\langle T\rangle$ and neglect
its fluctuation or variance $\langle\delta T^{2}\rangle=\langle T^{2}\rangle-\langle T\rangle^{2}$.
In this case, the contribution from the classical path $\langle T\rangle$
of $T$ dominates the path integral over $T$. By taking its average
value in the action, the partition function transforms into its semiclassical
result
\begin{equation}
Z=\int\mathscr{D}X\mathscr{D}Te^{iS[X,T]}\overset{S.C.}{\approx}\int\mathscr{D}Xe^{iS[X\left(\langle T\rangle\right)]}
\end{equation}
Then, up to an extra constant, we obtain an action that is highly
analogous to the action $S_{X}\left[X(\tau)\right]$, with the sole
difference being that the laboratory proper time parameter $\tau$
is replaced by the clock time $\langle T\rangle$, and the derivative
of $X$ with respect to $\tau$ in the kinetic term is now replaced
by a functional derivative of $X$ with respect to $T$
\begin{equation}
S[X\left(\langle T\rangle\right)]=\int dT\left[\frac{1}{2}M_{X}\left(\frac{\delta X}{\delta T}\right)^{2}-V(X)+\textrm{const}\right]
\end{equation}
in which the effective mass $M_{X}$ of the object with respect to
the clock time $T$ becomes
\begin{equation}
M_{X}=m_{X}\left\Vert \frac{d\tau}{d\langle T\rangle}\right\Vert \left(\frac{d\langle T\rangle}{d\tau}\right)^{2}=m_{X}\frac{d\langle T\rangle}{d\tau}
\end{equation}
Only when the clock runs at the same rate as the laboratory proper
time, that is, when the lapse function $\frac{d\langle T\rangle}{d\tau}=1$,
does $M_{X}$ equal $m_{X}$. In this sense, mass is not actually
an invariant under general coordinate transformations, but merely
an invariant under transformations between inertial coordinate systems. 

When the quantum fluctuations of the clock can be neglected (under
the semiclassical approximation of the clock), the position of the
clock's pointer is represented by a c-number rather than an operator.
In this case, the Schr\"{o}dinger equation governing evolution relative
to this clock can be derived as a semiclassical approximation to the
Wheeler-DeWitt equation. Therefore, the total Hamiltonian recovers
\begin{equation}
H[X\left(\langle T\rangle\right)]\overset{S.C.}{\approx}\frac{1}{2}M_{X}\left(\frac{dX}{dT}\right)^{2}+V(X)+\textrm{const}^{\prime}\label{eq:semi-classical Hamiltonian}
\end{equation}
The Hamiltonian $H_{X}+H_{T}$ in the Wheeler-DeWitt equation now
degenerates into the Hamiltonian $H[X\left(\langle T\rangle\right)]$
of the Schr\"{o}dinger equation with $\langle T\rangle$ as a parameter
time. Meanwhile, the solution to the Wheeler-DeWitt equation, the
entangled state $|X,T\rangle=\sum_{\tau}C_{\tau}|X(\tau)\rangle\otimes|T(\tau)\rangle$
reduces to the solution of the Schr\"{o}dinger equation, namely the
standard absolute quantum state $|X\left(\langle T\rangle\right)\rangle$.
Under the semiclassical approximation, approximating the coordinate
operator $T$ of the quantum clock as the parameter $\langle T\rangle$,
the quantum state $|X\left(\langle T\rangle\right)\rangle$ is equivalent
to projecting the entangled state $|X,T\rangle$ by using the condition
projector $\langle\tau^{\prime}|T\rangle=\sum_{\tau}A_{\tau}\delta_{\tau\tau^{\prime}}$,
which gives $\langle T|X,T\rangle$ (\ref{eq:projected entangled state}),
and $\delta_{\tau\tau^{\prime}}=\langle\tau^{\prime}|T(\tau)\rangle$
represents an infinitely precise, zero-width quantum clock eigenstate
$|T(\tau)\rangle$ under the semiclassical approximation. This yields
$|X\left(\langle T\rangle\right)\rangle=\sum_{\tau}\frac{C_{\tau}}{A_{\tau}}|X(\tau)\rangle$,
when the measuring instrument is in the reference state $|T\rangle=\sum_{\tau}A_{\tau}|T(\tau)\rangle$,
and hence the relative amplitude of the under-study system being in
the state $|X(\tau)\rangle$ is $\frac{C_{\tau}}{A_{\tau}}$.

Compared with the relative state approach, one of the advantages of
the functional integral approach is that the effects of the second-order
moment quantum fluctuation of the quantum clock $\langle\delta T^{2}\rangle$
(and that of the quantum reference frame fields) is more directly
to calculate in its perturbation expansion, which begins to have non-trivial
quantum effect in quantum gravity. Once the quantum fluctuations of
the clock, $\langle\delta T^{2}\rangle$, cannot be neglected---that
is, when the eigenstate of the clock, $\langle\tau^{\prime}|T(\tau)\rangle$,
can no longer be approximated semiclassically using a delta function
and instead requires consideration of its second-moment quantum fluctuation,
such as in the case of a coherent state with finite width---this
leads to numerous significant consequences. Firstly, although the
average energy of the system, $\langle E\rangle=0$, the fluctuations
of the clock, $\langle\delta T^{2}\rangle$, will result in vacuum
energy fluctuations, $\langle\delta E^{2}\rangle\neq0$, providing
the correct contribution of vacuum energy to the cosmological constant
\citep{Luo2014The,Luo2015Dark,Luo:2015pca,Luo:2019iby,Luo:2021zpi}.
Secondly, the Schr\"{o}dinger equation no longer strictly holds, which
causes the constant term in the semiclassical Hamiltonian (\ref{eq:semi-classical Hamiltonian})
to cease being a constant. This naturally introduces fluctuation effects
akin to some ``inertial forces'', as demonstrated in \citep{Anandan:1990fq,2024arXiv240809630L}.

When this model of a quantum clock with a single degree of freedom
is generalized to a quantum spacetime reference frame with multiple
degrees of freedom, the aforementioned results are extended in a similar
manner: the action $S_{T}$ of the single-degree-of-freedom quantum
clock is generalized to a non-linear sigma model with 4-degrees of
freedom $X,Y,Z,T$, serving as the quantum spacetime reference frame
fields having a vanished Hamiltonian. Meanwhile, the action $S_{X}$
of the object's position can be generalized to the action of a general
quantum field, e.g. $\phi$, representing the quantum system under-study.
The global parameter $\tau$ is extended from a single parameter to
the four Minkowski coordinate parameters $x,y,z,t$ of the laboratory
spacetime, which act as the base space of the non-linear sigma model,
while the target space corresponds to the actual spacetime $X,Y,Z,T$.
The resulting model, under the semiclassical approximation, yields
the effective action of the quantum field $\phi$ on a general curved
spacetime background. Furthermore, when considering the effects of
second-order quantum fluctuations of the quantum spacetime reference
frame fields beyond the semiclassical approximation, the effects of
``inertial forces''---namely, gravity---automatically emerge,
and, under certain approximation, it takes on the form of Einstein
gravity \citep{Luo:2021zpi,2024arXiv240809630L,Luo:2025cer}.

\bibliographystyle{unsrt}

\end{document}